\documentclass{rsproca_new}
\jname{rspa}

\usepackage{tikz,xcolor,hyperref}

\definecolor{lime}{HTML}{A6CE39}
\DeclareRobustCommand{\orcidicon}{%
	\begin{tikzpicture}
	\draw[lime, fill=lime] (0,0) 
	circle [radius=0.16] 
	node[white] {{\fontfamily{qag}\selectfont \tiny ID}};
	\draw[white, fill=white] (-0.0625,0.095) 
	circle [radius=0.007];
	\end{tikzpicture}
	\hspace{-2mm}
}

\foreach \x in {A, ..., Z}{%
	\expandafter\xdef\csname orcid\x\endcsname{\noexpand\href{https://orcid.org/\csname orcidauthor\x\endcsname}{\noexpand\orcidicon}}
}

\begin{document}

\title{Acoustic and inertial modes in planetary-like rotating ellipsoids}

\author{J\'er\'emie Vidal$^{1}$ and David C\'ebron$^{2}$}

\address{$^{1}$Department of Applied Mathematics, University of Leeds, LS2 9JT, UK\\
$^{2}$Universit\'e Grenoble Alpes, CNRS, ISTerre, France\\
\orcidA{ JV, \href{https://orcid.org/0000-0002-3654-6633}{0000-0002-3654-6633}}\\
}

\subject{geophysics, wave motion, fluid mechanics}


\corres{J\'er\'emie Vidal\\
\email{vidalje63@gmail.com}}
\esm{}

\begin{abstract}
The bounded oscillations of rotating fluid-filled ellipsoids can provide physical insight into the flow dynamics of deformed planetary interiors. 
The inertial modes, sustained by the Coriolis force, are ubiquitous in rapidly rotating fluids and Vantieghem (2014, \emph{Proc. R. Soc. A}, \textbf{470}, 20140093, \href{https://dx.doi.org/10.1098/rspa.2014.0093}{doi:10.1098/rspa.2014.0093}) pioneered a method to compute them n incompressible fluid ellipsoids.
Yet, taking density (and pressure) variations into account is required for accurate planetary applications, which has hitherto been largely overlooked in ellipsoidal models. 
To go beyond the incompressible theory, we present a Galerkin method in rigid coreless ellipsoids, based on a global polynomial description.  
We apply the method to investigate the normal modes of fully compressible, rotating and diffusionless fluids. 
We consider an idealized model, which fairly reproduces the density variations in the Earth's liquid core and Jupiter-like gaseous planets. 
We successfully benchmark the results against standard finite-element computations. 
Notably, we find that the quasi-geostrophic inertial modes can be significantly modified by compressibility, even in moderately compressible interiors.
Finally, we discuss the use of the normal modes to build reduced dynamical models of planetary flows. 
\end{abstract}


\begin{fmtext}
\section{Introduction}
Modelling the flow dynamics of rapidly rotating planetary interiors is a very challenging task in fluid mechanics.  
Physical insight can be gained from direct numerical simulations (DNS) but, despite many computational improvements, even the most demanding DNS \cite{schaeffer2017turbulent,sheyko2018scale} operate for parameters far from the planetary range. 
Thus, our physical understanding of rotating planetary flows remains largely incomplete. 
\end{fmtext}


\maketitle

Using model reduction can fortunately provide a complementary knowledge. 
The theory consists in reducing the primitive equations into a lower-dimensional (dynamical) problem. 
Several data-driven methods have been developed in fluid mechanics (e.g. the dynamic mode decomposition), but physically based approaches can be also employed. 
For instance, insightful asymptotic models have been developed in spheres to study rapidly rotating convection \cite{guervilly2019nature} or dynamo magnetic fields \cite{chen2018optimal,daria2019dynamo}. 
More generally, the normal modes could be used to incorporate the key characteristics of the system \cite[for planetary flows]{kong2018origin,kloss2019time}. 

A striking feature of rapidly rotating fluids is the ubiquitous presence of inertial modes, which are sustained by the Coriolis force and have been mainly studied for incompressible fluids \cite{greenspan1968theory}.
They are governed by a hyperbolic operator, leading to an ill-posed problem when associated with boundary conditions \cite{rieutord2000wave}. 
Pathological modes were soon conjectured \cite{stewartson1969pathological}, and indeed found in spherical shells \cite{rieutord2001inertial,rieutord2018axisymmetric}. 
However, the inviscid incompressible modes are smooth in full spheres and ellipsoids \cite{backus2017completeness,ivers2017enumeration}, and have been even used to initiate a consistent model of rapidly rotating flows \cite{zhang2017theory}.
The validity of such a model remains elusive for compressible planetary interiors, subject to density stratification, and the fully compressible inertial modes have received scant theoretical consideration so far  \cite{papaloizou1978non,lockitch1999r,ivanov2010inertial}. 
For simplicity, we could use sound-proof formulations of the compressible equations (e.g. the anelastic approximation), but their applicability is still debated \cite{dintrans2001comparison,wood2016oscillatory,verhoeven2018validity}. 
Thus, an unambiguous fully compressible theory of the inertial modes in the presence of planetary-like density (and pressure) variations is desirable. 

Another limitation in planetary models is to consider (weakly) deformed spherical geometries. 
Planets are indeed rather ellipsoidal at leading order, for instance due to centrifugal effects \cite{zhang2017shape}, orbital forcings \cite{le2015flows}, or mantle convection \cite{davies2014strength}. 
Even if the boundary deformation is often small, it can be responsible for important physical mechanisms.
Flow instabilities that involve triadic interactions of inertial modes in ellipsoidal domains can sustain large-scale magnetic fields \cite{cebron2014tidally,reddy2018turbulent,vidal2018magnetic}, contrary to single inertial modes \cite{herreman2011stokes}. 
The ellipsoid is thus worth considering (as a first step), but solving the problem is hampered by the mathematical complexity of the geometry \cite{bryan1889waves}. 
Faced with this difficulty, near-spherical boundaries could be described using perturbation theory (e.g. \cite{rekier2018inertial} for incompressible fluids) or non-orthogonal mappings of the coordinates (e.g. \cite{reese2006acoustic,seyed2014dynamics} for axisymmetric compressible bodies), but numerical convergence could be difficult to achieve with low-dimensional models (e.g. \cite{seyed2007inertial} for the spherical inertial modes). 

Consequently, the next theoretical step is to find a suitable mathematical description of the planetary compressible modes in the coreless ellipsoid. 
The compressible modes are indeed expected to be regular in the latter geometry, to agree with the anelastic (inertial) modes \cite{busse2005slow,wu2005origin,clausen2014elliptical} or the non-rotating acoustic modes in thermally stratified spheres \cite{koulakis2018acoustic}.
In the incompressible regime, Vantieghem \cite{vantieghem2014inertial} pioneered a groundbreaking method to compute the inertial modes in rigid ellipsoids, based on global polynomial elements in the Cartesian coordinates. 
Vidal \emph{et al.} \cite{vidal2020acoustics} recently followed the same path, and devised an admissible polynomial description of uniform-density compressible flows in rigid ellipsoids. 
Although polynomial descriptions may appear effective only when the problem is relatively simple, they can compensate for this by providing better physical insight into the problem. 
Moreover, polynomial methods in ellipsoids are not restricted in practice to uniform-density fluids, as pointed out by Chandrasekhar in his seminal monograph \cite[see the epilogue]{chandrasekhar1969ellipsoidal}.

The present study generalizes the theoretical works on incompressible \cite{vantieghem2014inertial} and uniform-density compressible fluids \cite{vidal2020acoustics}, to take density (and pressure) variations into account within an idealised ellipsoidal model of fully compressible (heterogeneous) planetary interiors. 
Beyond the theoretical motivation, such a model could be used as a preparatory step to benchmark (or develop) more complicated models of planets.  
The paper is organized as follows. 
We describe the compressible model in \S\ref{sec:model}, and we present the polynomial method in \S\ref{sec:modalpb}. 
Numerical results are presented in \S\ref{sec:numerics}, and some planetary implications are discussed in \S\ref{sec:discuss}. 
We end the paper with concluding remarks in \S\ref{sec:ccl}.

\newpage
\section{Formulation of the problem}
\label{sec:model}
\subsection{Linearized compressible equations}
Throughout the paper, we employ either dimensional or dimensionless variables. 
For clarity, we denote the dimensional variables that admit dimensionless counterparts with the superscript ${}^\ast$. 
We consider a fluid-filled (triaxial) ellipsoid of arbitrary semi-axes $[a,b,c]$ and volume $V$. 
The ellipsoidal cavity is co-rotating with the fluid at the angular velocity $\boldsymbol{\Omega}^\ast = \Omega_s \, \boldsymbol{1}_z$, where $ \boldsymbol{1}_z$ is the unit vector along the $z$-axis. 
In the following, we work exclusively in the co-rotating frame, where the ellipsoidal boundary $\partial V$ is stationary. 
We employ the Cartesian coordinates $(x,y,z)$, and introduce the position vector $\boldsymbol{r}^\ast$. 
In the rotating frame, the ellipsoidal boundary $\partial V$ is described by the Cartesian equation $F(x,y,z)= 1$ with the shape function $F(x,y,z)=({x}/{a})^2 + ({y}/{b})^2 + ({z}/{c})^2$. 
We expand the velocity, the density and the pressure as small perturbations $[\boldsymbol{u}_1^\ast, \rho_1^\ast, p_1^\ast]$ around a steady and motionless background state, and we neglect the perturbation of the gravitational potential for simplicity (Cowling approximation \cite{cowling1941non}). 
The reference state is characterized by the background density $\rho_0^\ast (\boldsymbol{r}^\ast)$, the background pressure $P_0^\ast (\boldsymbol{r}^\ast)$, the adiabatic speed of sound $C_0^\ast (\boldsymbol{r}^\ast)$, and the gravity field $\boldsymbol{g}^\ast (\boldsymbol{r}^\ast)$. 

Small viscous effects are expected rapidly rotating interiors, as measured by the Ekman number $Ek = \nu/(\Omega_s a^2)$ where $\nu$ is the (laminar) kinematic viscosity. 
The wave dynamics we are considering operates on time scales much shorter than the spin-up time $Ek^{-1/2} \Omega_s^{-1}$ \cite{greenspan1968theory}, or the viscous time $(Ek \, \Omega_s)^{-1}$. 
Indeed, typical planetary values are $Ek \sim 10^{-13}$ in liquid cores of Galilean moons \cite{cebron2012elliptical}, $Ek \sim 10^{-15}$ in the Earth's liquid core, and $Ek \sim 10^{-18} - 10^{-16}$ for Jovian planets. 
On short time scales, viscosity is mainly responsible for small viscous damping of the modes (due to viscous effects at the boundary \cite{abney1996ekman,glampedakis2006ekman}), and the inviscid modes could be even excited in the bulk in the presence of non-vanishing viscosity
(by analogy with the incompressible modes \cite{aldridge1969axisymmetric}). 
Since viscous effects appear negligible at leading order, we focus on non-viscous fluids.
Similarly, we neglect thermal diffusion and consider isentropic perturbations.

The diffusionless perturbations are thus given by the linearized compressible equations
\begin{subequations}
\label{eq:compressible1}
\allowdisplaybreaks
\begin{align}
	\rho_0^\ast \left ( \frac{\partial \boldsymbol{u}_1^\ast}{\partial t^\ast} + 2 \, \boldsymbol{\Omega}^\ast \times \boldsymbol{u}_1^\ast \right ) &= -\boldsymbol{\nabla} p_1^\ast + \rho_1^\ast \, \left ( \boldsymbol{g}^\ast + \boldsymbol{g}_c^\ast \right ), \label{eq:momentum1} \\
	\frac{\partial \rho_1^\ast}{\partial t^\ast} + \boldsymbol{\nabla} \boldsymbol{\cdot} (\rho_0^\ast \boldsymbol{u}_1^\ast) &= 0, \label{eq:density1}
\end{align}
\end{subequations}
with the centrifugal gravity $\boldsymbol{g}_c^\ast$. 
For isentropic perturbations, the entropy equation reduces to \cite[see equation (43)]{roberts1967introduction}
\begin{equation}
    \frac{\partial p_1^\ast}{\partial t^\ast} + (\boldsymbol{u}_1^\ast \boldsymbol{\cdot} \boldsymbol{\nabla}) \, P_0^\ast = {C_0^\ast}^2 \left ( \frac{\partial \rho_1^\ast}{\partial t^\ast} + (\boldsymbol{u}_1^\ast \boldsymbol{\cdot} \boldsymbol{\nabla}) \, \rho_0^\ast \right ) =  - \rho_0^\ast {C_0^\ast}^2 \, \boldsymbol{\nabla} \boldsymbol{\cdot} \boldsymbol{u}_1^\ast,
    \label{eq:pressure1}
\end{equation}
where we have used mass equation (\ref{eq:density1}) in the last equality. 
Finally, equations (\ref{eq:compressible1}) are supplemented with boundary conditions for the velocity. 
Motivated by planetary liquid cores that are surrounded by solid mantles, we consider an impenetrable (rigid) stationary boundary on which the velocity field satisfies the non-penetration condition $\boldsymbol{u}_1^\ast \boldsymbol{\cdot} \boldsymbol{1}_n = 0$, where $\boldsymbol{1}_n$ is the unit vector normal to the boundary. 
The theory does not require any additional boundary conditions in the diffusionless regime. 
Note that a free-surface boundary condition would be more appropriate for gas giants (or stellar envelopes). 
However, free-surface effects are expected to be of second-order importance for the interior motions (at least for the incompressible modes \cite{braviner2014tidal}, but it is still disputed with compressibility \cite{goodman2009dynamical}), such that impenetrable boundaries are often considered in numerical models of fully compressible spherical convection  \cite{kapyla2017convection,liu2019onset}.

\subsection{Planetary range of parameters}
\label{subsec:planets}
\begin{table}
    \caption{Planetary parameters. Equatorial ellipticity $\beta=|a^2-b^2|/(a^2+b^2)$. Polar flattening $\epsilon = 1-(c/a)$. Density contrast $H_\rho = -\log(1-\alpha)$. Rotational Mach number $M_\Omega=a \Omega_s/C_c$.
    References: \cite{zhang2017shape,labrosse2015thermal,cebron2012elliptical,evonuk2012simulating}.}
    \label{table:planets}
    \centering
    \begin{tabular}{lccccc}
    \hline
        & $\beta$ & $\epsilon$ & $H_\rho$ & $\alpha$ & $M_\Omega$ \\ [1ex]
    Planet's core (Earth)  & $\leq 10^{-3}$ & $2.5 \times 10^{-4}$ & $2.3\times 10^{-1}$ & $2.1 \times 10^{-1}$ & $0.02$ \\ 
    Moon's core (Ganymede) & $3.7 \times 10^{-4}$ & $5 \times 10^{-4}$ & $3.0\times 10^{-2}$ & $2.9 \times 10^{-2}$ & $ 10^{-3}$ \\ 
    Gas giant (Jupiter)  & $1.6 \times 10^{-7}$ & $6.3 \times 10^{-2}$ & $4.6$ & $9.9 \times 10^{-1}$ & $0.3$ \\ 
    \hline
    \end{tabular}
\end{table}

We will mainly employ dimensionless units (for numerical convenience), writing the dimensionless variables without the superscript ${}^\ast$. 
We choose the semi-axis $a$ as length scale, the central speed of sound $C_c$ (i.e. at $x=y=z=0$) as velocity scale, $a/C_c$ as time scale, the central density $\rho_c$ as density scale, $\rho_c {C_c}^2$ as pressure scale and $C_c^2/a$ as gravity scale. 
Typical planetary estimates are summarized in table \ref{table:planets}. 

The departure from the spherical geometry 
is measured by the equatorial ellipticity $\beta= |a^2-b^2|/(a^2+b^2)$, and the polar flattening $\epsilon = 1-(c/a)$. 
The equatorial deformation of planetary liquid cores is small, typically $10^{-7} \leq \beta \leq 10^{-3}$ in the Earth \cite{gerick2020mhd}.
So, we expect the equatorial ellipticity to have small effects on the modal frequencies (as reported in uniform-density ellipsoids \cite{vantieghem2014inertial}), but small ellipticity effects can be important to favour couplings between the normal modes (this is beyond the scope of the study). 
The polar flattening is usually much larger, especially in rapidly rotating gas giants. 
Centrifugal effects cause indeed strong departures from sphericity, with for instance $\epsilon \simeq 0.06$ in Jupiter \cite{zhang2017shape}. 

Independently of the centrifugal deformation, we introduce the dimensionless rotational Mach number $M_\Omega = {a \Omega_s}/{C_c}$, which compares the rotational and sonic time scales. 
Rotating planets are characterized by moderately small values of $M_\Omega$, typically $M_\Omega \sim 0.02 - 0.03$ in the Earth's liquid core \cite{labrosse2015thermal}, but gas giants have larger values (e.g. $M_\Omega \sim 0.3$ for Jupiter \cite{zhang2017shape}).

Planetary interiors are also characterized by the density contrast $H_\rho = \log (\rho_c/\rho_b)$ between the centre and the outer boundary, where $\rho_b$ is the density at the outer boundary. 
Typical values are $H_\rho = \mathcal{O}(10^{-2})$ in the liquid cores of Galilean moons \cite{evonuk2012simulating}, $H_\rho = \mathcal{O}(10^{-1})$ in planetary liquid cores \cite{labrosse2015thermal}, and $H_\rho > 1$ in the interiors of Jovian planets (excluding the outermost atmospheres). 

\subsection{Reference state}
We describe the background reference state $[\rho_0^\ast, P_0^\ast, C_0^\ast, \boldsymbol{g}^\ast]$. 
Planetary fluid interiors are often convectively unstable but usually with a low degree of super-adiabaticity, and so we assume neutrally buoyant (isentropic) interiors at leading order. 
The reference state is thus given by the hydrostatic equilibrium and the isentropic equation of state
\begin{subequations}
\label{eq:hydrostaticeq}
\begin{equation}
    \boldsymbol{\nabla} P_0^\ast = \rho_0^\ast \, \left ( \boldsymbol{g}^\ast + \boldsymbol{g}_c^\ast \right ), \quad {C_0^\ast}^2 \, \boldsymbol{\nabla} \rho_0^\ast = \rho_0^\ast \,  \left ( \boldsymbol{g}^\ast + \boldsymbol{g}_c^\ast \right ).
    \tag{\theequation a,b}
\end{equation}
\end{subequations}
The gravity $\boldsymbol{g}^\ast$ has the typical amplitude $C_c^2/a$, whereas the centrifugal gravity $\boldsymbol{g}_c^\ast = -\, \boldsymbol{\Omega}^\ast \times (\boldsymbol{\Omega}^\ast \times \boldsymbol{r}^*)$ has the typical magnitude $\Omega_s^2 a$. 
The centrifugal gravity is smaller than $\boldsymbol{g}^\ast$ as long as $M_\Omega^2 \ll 1$ in dimensionless units, as often found in planetary interiors (table \ref{table:planets}). 
Since we aim to model arbitrary ellipsoidal boundaries, we discard the centrifugal gravity in (\ref{eq:hydrostaticeq}). 
This corresponds to the most general situation, where the boundary can be deformed by mechanisms that are independent of the rotation rate of the fluid (e.g. tides or mantle convection). 
Moreover, it will allow us to disentangle carefully the effects of rotation and ellipticity on the modes (contrary to \cite{reese2006acoustic} for the acoustic modes).
Yet, centrifugally distorted boundaries \cite{chandrasekhar1969ellipsoidal} can still be modelled by considering the appropriate ellipsoidal axes.

For mathematical simplicity, we also seek an idealized model consistent with (\ref{eq:hydrostaticeq}) 
that continuously varies from Earth-like to Jovian models. 
Planetary liquid cores have indeed non-vanishing density and pressure fields on the boundary (contrary to gas giants). 
Hence, we consider density and pressure profiles in the form
\begin{subequations}
\label{eq:backgroundrho0P0}
\begin{equation}
    \rho_0^\ast =  \rho_c \left [ 1 - \alpha F \right ] \quad \text{and} \quad P_0^\ast =  P_c \left [ 1 - \alpha F \right ]^2, 
    \tag{\theequation a,b}
\end{equation}
\end{subequations}
with the central density $\rho_c$, the central pressure $P_c$, and $0 \leq \alpha \leq 1$ an adjustable parameter.
The latter coefficient allows us to continuously describe planetary liquid cores (with a non-zero density on the boundary when $\alpha \neq 1$) and Jovian planets (for which $\rho_0\to0$ at the boundary, i.e. $\alpha \to 1$), whereas the problem reduces to uniform-density compressible fluids with $\alpha = 0$ \cite{vidal2020acoustics}. 
Moreover, the definition of the density contrast translates into $H_\rho =  -\log (1-\alpha)$ with this model, and the planetary values for $\alpha$ are given in table \ref{table:planets}. 
Then, hydrostatic equilibrium (\ref{eq:hydrostaticeq}) gives the gravity field and the speed of sound
\begin{subequations}
\begin{align}
    \boldsymbol{g}^\ast &= - \alpha C_c^2 \, \boldsymbol{\nabla} F = -2 \alpha C_c^2 \left ( \frac{x}{a^2}  \boldsymbol{1}_x + \frac{y}{b^2} \boldsymbol{1}_y + \frac{z}{c^2} \boldsymbol{1}_z \right ),
    \label{eq:backgroundg0} \\
    C_0^\ast &=  C_c \, \sqrt{1 - \alpha F}, \label{eq:backgroundC0}
\end{align}
\end{subequations}
with $C_c = \sqrt{{2 P_c}/{\rho_c}}$ the speed of sound at the centre, and $[\boldsymbol{1}_x, \boldsymbol{1}_y, \boldsymbol{1}_z]$ the unit Cartesian vectors. 
More realistic polynomial solutions of (\ref{eq:hydrostaticeq}) could be considered, see the  electronic supplementary material. 

\begin{figure}
    \centering
    \begin{tabular}{cc}
         \includegraphics[width=0.45\textwidth]{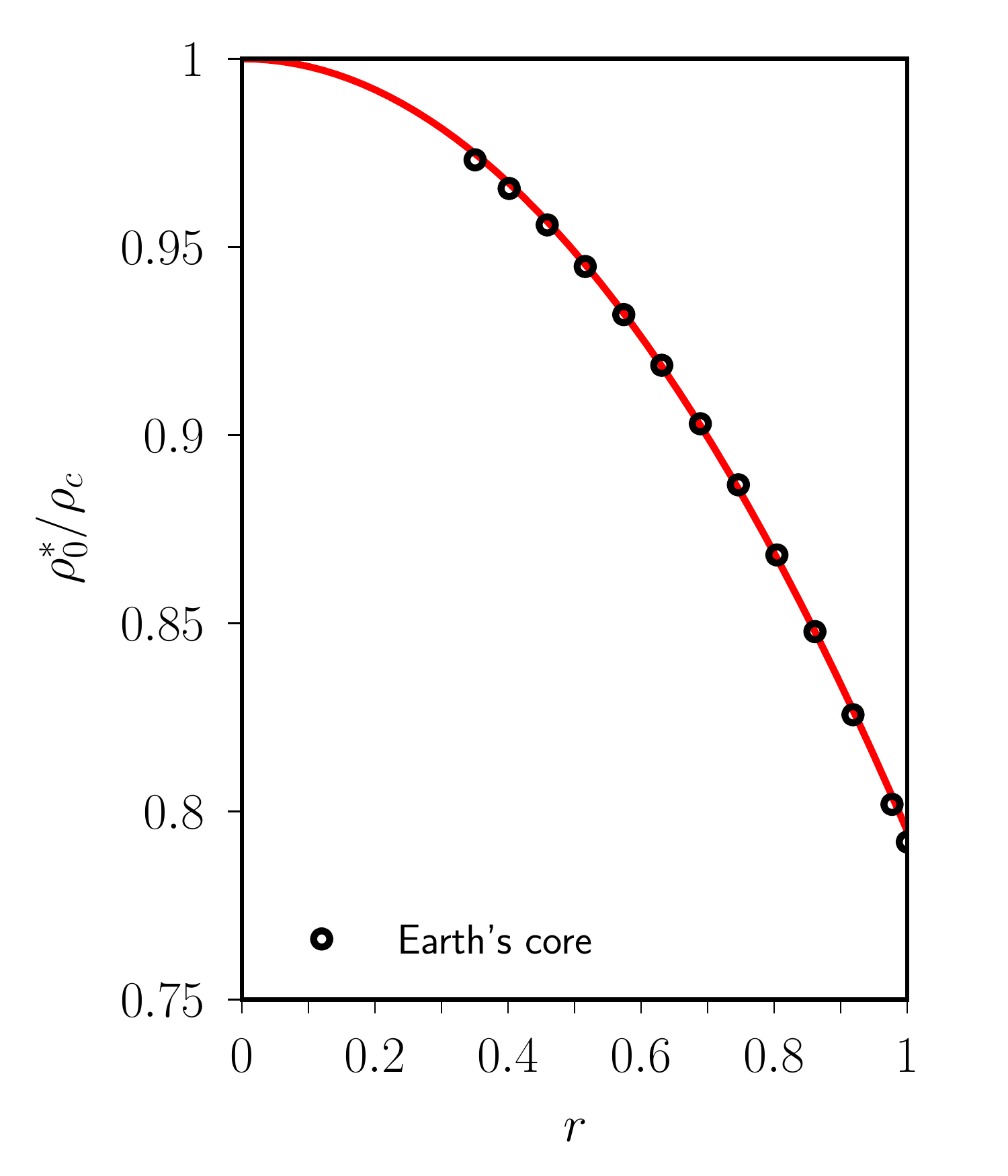} &
         \includegraphics[width=0.45\textwidth]{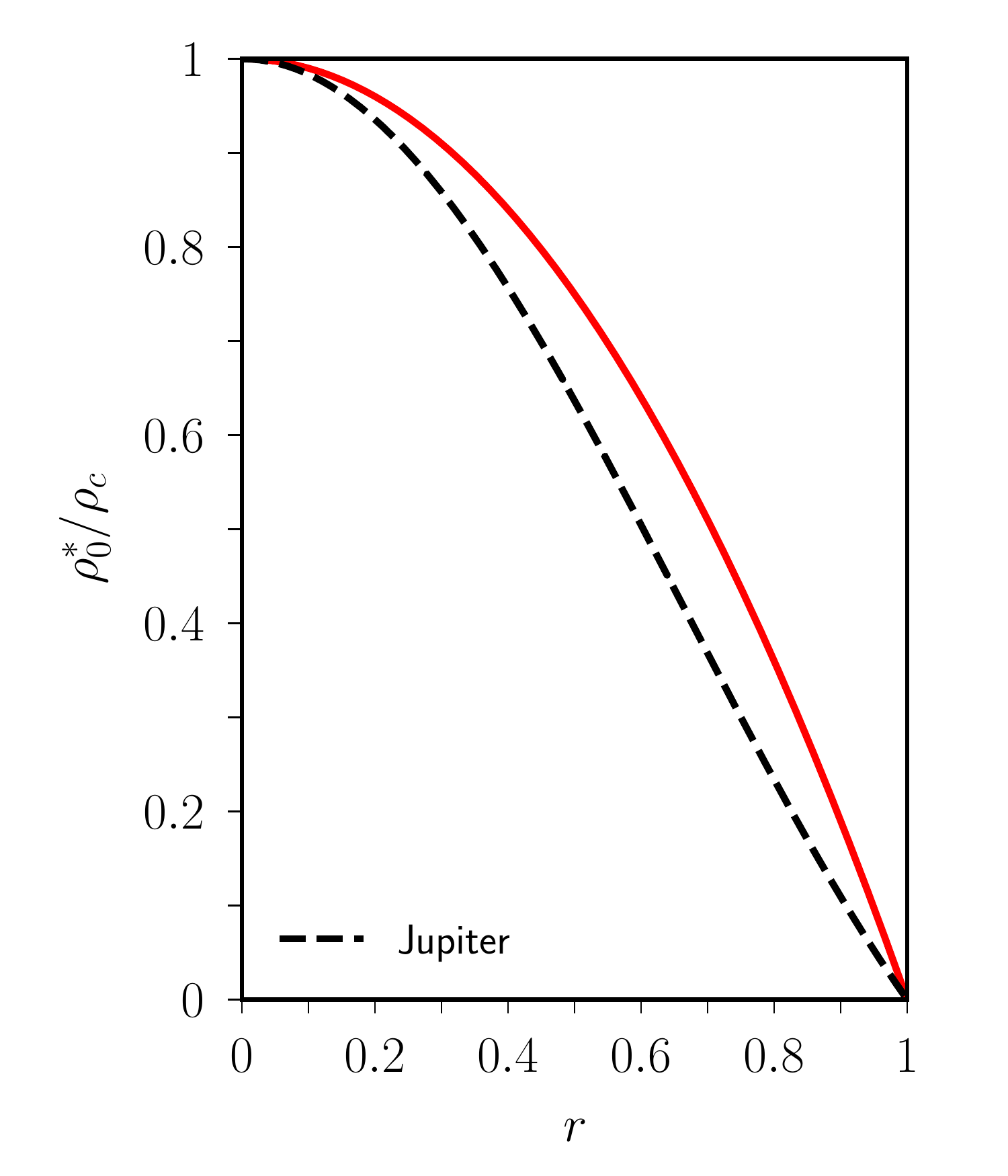} \\
           $\alpha=0.205$ & $\alpha=1$ \\
    \end{tabular}
    \caption{Dimensionless density profiles (solid red lines), as a function of the dimensionless radius $r$ for a spherical model. Open circles: Earth's liquid core \cite{labrosse2015thermal}. Dashed curve: leading-order Jupiter's profile \cite{chandrasekhar1958introduction}. (Online version in colour.)
    }
    \label{fig:refstate}
\end{figure}

Profiles (\ref{eq:backgroundrho0P0}) to (\ref{eq:backgroundC0}) are mathematically simple, but are at the same time reasonably realistic at leading order for planetary interiors.  
They decrease monotonically with the radial-like distance $F^{1/2}$, and have constant density-pressure values on each ellipsoidal surface (in agreement with the theory of compressible figures of equilibrium \cite{lai1993ellipsoidal}). 
Moreover, profiles (\ref{eq:backgroundrho0P0}a) fairly accommodate the expected density variations of planetary interiors, as shown in figure \ref{fig:refstate}.
We have superimposed realistic profiles for the Earth's liquid core \cite{labrosse2015thermal} and Jupiter \cite{chandrasekhar1958introduction}. 
A very good agreement is found for the Earth-like model. 
Likewise, profile (\ref{eq:backgroundrho0P0}a) with $\alpha=1$ does not deviate by more than a few percent from the leading-order component of a realistic Jovian profile. 
The background state is further discussed in the electronic supplementary material.

\subsection{Dimensionless wave-like equation}
\label{subsec:adimequations}
To simplify the mathematical analysis, we combine equations (\ref{eq:compressible1}) to obtain a master wave-like equation for the (Lagrangian) displacement vector $\boldsymbol{\xi}_1$, defined by \cite{chandrasekhar1969ellipsoidal} 
\begin{subequations}
\label{eq:lagrangiandef}
\begin{equation}
    \boldsymbol{u}_1 = \partial \boldsymbol{\xi}_1 /\partial t, \quad 
    p_1 = - \rho_0 {C_0}^2 \, \boldsymbol{\nabla} \boldsymbol{\cdot} \boldsymbol{\xi}_1  - (\boldsymbol{\xi}_1 \boldsymbol{\cdot} \boldsymbol{\nabla}) \, P_0, 
    \quad \rho_1 = - \boldsymbol{\nabla} \boldsymbol{\cdot} (\rho_0 \boldsymbol{\xi}_1).
    \tag{\theequation a--c}
\end{equation}
\end{subequations}
Momentum equation (\ref{eq:momentum1}) can be now expressed solely as a function of $\boldsymbol{\xi}_1$ such that
\begin{equation}
    \rho_0 \left (\frac{\partial^2 \boldsymbol{\xi}_1}{\partial t^2} + 2 \, M_\Omega \, \boldsymbol{1}_z \times \frac{\partial\boldsymbol{\xi}_1}{\partial t} \right ) = \boldsymbol{\nabla} \left [ \rho_0 {C_0}^2 \, \boldsymbol{\nabla} \boldsymbol{\cdot} \boldsymbol{\xi}_1  + (\boldsymbol{\xi}_1 \boldsymbol{\cdot} \boldsymbol{\nabla}) \, P_0 \right ] - \boldsymbol{\nabla} \boldsymbol{\cdot} (\rho_0 \boldsymbol{\xi}_1) \, \boldsymbol{g}, 
    \label{eq:compressible2}
\end{equation}
with $\boldsymbol{g} = (a/C_c^2) \, \boldsymbol{g}^\ast$. 
We have neglected the centrifugal gravity in (\ref{eq:compressible2}) because the corresponding buoyancy term, of dimensionless amplitude $M_\Omega^2$, is smaller than the other terms in the planetary range $M_\Omega \leq 1$. 
Then, we simplify the right-hand side of (\ref{eq:compressible2}) to obtain (after little algebra)
\begin{equation}
     \rho_0 \left (\frac{\partial^2 \boldsymbol{\xi}_1}{\partial t^2} + 2 M_\Omega \, \boldsymbol{1}_z \times \frac{\partial \boldsymbol{\xi}_1}{\partial t} \right ) = \boldsymbol{\nabla} \left [ C_0^2 \, \boldsymbol{\nabla} \boldsymbol{\cdot} (\rho_0 \boldsymbol{\xi}_1) \right ] - \boldsymbol{\nabla} \boldsymbol{\cdot} (\rho_0 \boldsymbol{\xi}_1) \, \boldsymbol{g}
    \label{eq:compressible3}
\end{equation}
by virtue of hydrostatic equilibrium (\ref{eq:hydrostaticeq}). 
Equations (\ref{eq:compressible2})-(\ref{eq:compressible3}) are the governing equations of the rotating compressible modes in the diffusionless theory (for any isentropic reference states).
They are supplemented here with the non-penetration condition $\boldsymbol{\xi}_1 \boldsymbol{\cdot} \boldsymbol{1}_n = 0$ on the rigid boundary. 

\section{Quadratic eigenvalue problem}
\label{sec:modalpb}
\subsection{Infinite-dimensional formulation}
\label{subsec:infinite}
We seek diffusionless modal solutions upon our isentropic reference state as
\begin{equation}
    \boldsymbol{\xi}_1 (\boldsymbol{r},t) = \boldsymbol{\zeta} (\boldsymbol{r}) \exp(\lambda t), \quad \boldsymbol{\zeta} \boldsymbol{\cdot} \boldsymbol{1}_n = 0 \ \, \text{on} \ \, \partial V,
    \label{eq:modalansatz}
\end{equation}
with $\lambda \in \mathbb{C}$ the eigenvalue and $\boldsymbol{\zeta} (\boldsymbol{r})$ the complex-valued spatial dependence. 
We substitute expansion (\ref{eq:modalansatz}) into wave-like equation (\ref{eq:compressible3}) to get the quadratic eigenvalue problem (QEP)
\begin{equation}
   \lambda^2 \boldsymbol{\zeta} + \lambda \, \boldsymbol{\mathcal{C}} ( \boldsymbol{\zeta}) + \boldsymbol{\mathcal{K}} (\boldsymbol{\zeta}) = \boldsymbol{0},
   \label{eq:compressibleQEP}
\end{equation}
with the two linear operators
\begin{subequations}
\label{eq:operatorsA2A1A0}
\begin{equation}
   \boldsymbol{\mathcal{C}}(\boldsymbol{\zeta}) = 2 M_\Omega \, \boldsymbol{1}_z \times  \boldsymbol{\zeta}, \quad 
   \rho_0 \, \boldsymbol{\mathcal{K}} (\boldsymbol{\zeta}) = -\boldsymbol{\nabla} \left [ C_0^2 \, \boldsymbol{\nabla} \boldsymbol{\cdot} (\rho_0 \boldsymbol{\zeta})  \right ] + \boldsymbol{\nabla} \boldsymbol{\cdot} (\rho_0 \boldsymbol{\zeta}) \, \boldsymbol{g}.
    \tag{\theequation a,b}
\end{equation}
\end{subequations}
To determine the symmetries of the problem, we define the weighted inner product between two complex-valued vector fields  $[\boldsymbol{a},\boldsymbol{b}]$
\begin{equation}
    \langle \boldsymbol{a}, \boldsymbol{b} \rangle_{\rho_0} = \int_V \rho_0 \, \boldsymbol{a}^{\dagger} \boldsymbol{\cdot} \boldsymbol{b} \ \mathrm{d} V,
    \label{eq:scalprod0}
\end{equation}
where ${}^{\dagger}$ denotes the complex conjugate (as in \cite{vantieghem2014inertial}). 
The Coriolis operator $\boldsymbol{\mathcal{C}}$ is skew-adjoint with respect to (\ref{eq:scalprod0}) \cite{lynden1967stability}, whereas the elastic-gravitational operator $\boldsymbol{\mathcal{K}}$ is self-adjoint \cite{valette1989spectre}. 

From the symmetries of the operators, if $[\lambda, \boldsymbol{\zeta}]$ is a solution of QEP (\ref{eq:compressibleQEP}), then $[\lambda^{\dagger}, \boldsymbol{\zeta}^{\dagger}]$ is also a solution \cite{lynden1967stability}. 
Moreover, QEP (\ref{eq:compressibleQEP}) can possess unstable modes (with $\Re_e (\lambda) \geq 0$) only if some modal solutions are unstable when $\boldsymbol{\mathcal{C}} = \boldsymbol{0}$ \cite[see section II]{barston1967eigenvalue}. 
Since our background reference state is neutrally stable, diffusionless and motionless, we conclude that the (diffusionless) modes with $\boldsymbol{\mathcal{C}} = \boldsymbol{0}$  are stable.
Thus, the QEP admits only stable eigenvalues, and we denote $\Im_m (\lambda) = \omega$ the real-valued angular frequency in the following.

\subsection{Spectral decomposition}
We consider solutions with finite kinetic energies $\langle \boldsymbol{\zeta}, \boldsymbol{\zeta} \rangle_{\rho_0} < \infty$, and seek a spectral decomposition in rigid ellipsoids to achieve numerical convergence in the presence of density (and pressure) variations.
We start with the general weighted Helmholtz decomposition
\begin{equation}
    \boldsymbol{\zeta} = (1/\rho_0) \, \boldsymbol{\nabla} \times \boldsymbol{\Psi} + \boldsymbol{\nabla} \Phi, \quad \boldsymbol{\zeta} \boldsymbol{\cdot} \boldsymbol{1}_n = 0 \ \,  \text{on} \ \, \partial V,
    \label{eq:helmholtzweighted}
\end{equation}
which is restricted here to displacements satisfying the non-penetration boundary condition (contrary to \cite{lebovitz1989stability}). 
The vector space of the solutions is divided into two sub-spaces that are mutually orthogonal since, with respect to inner product (\ref{eq:scalprod0}), we have
\begin{equation}
    \langle (1/\rho_0) \, \boldsymbol{\nabla} \times \boldsymbol{\Psi}, \boldsymbol{\nabla} \Phi \rangle_{\rho_0} = \int_V (\boldsymbol{\nabla} \times \boldsymbol{\Psi}^{\dagger}) \boldsymbol{\cdot} \boldsymbol{\nabla} \Phi \, \mathrm{d} V = 0.
    \label{eq:weighedinnerprodcuct}
\end{equation}
The weighted Helmholtz decomposition has proven valuable to determine the invariant sub-spaces of the normal modes \cite{sobouti1981potentials}.
Yet, despite its theoretical advantages, decomposition (\ref{eq:helmholtzweighted}) must be tailored to the background configuration (because of its dependence on $\rho_0$). 
This could be awkward for the numerical analysis (in order to survey the parameter space). 

We describe instead $\rho_0 \boldsymbol{\zeta}$ with the Helmholtz decomposition in rigid ellipsoids
\begin{equation}
    \rho_0 \boldsymbol{\zeta} = \boldsymbol{\nabla} \times \widehat{\boldsymbol{\Psi}} + \boldsymbol{\nabla} \widehat{\Phi}, \quad \boldsymbol{\zeta} \boldsymbol{\cdot} \boldsymbol{1}_n = 0 \ \,  \text{on} \ \, \partial V,
    \label{eq:helmholtzweighted2}
\end{equation}
which is actually equivalent to the weighted decomposition (see appendix \ref{appendix:spectral}). 
Decomposition (\ref{eq:helmholtzweighted2}) involves the two sub-spaces 
\begin{subequations}
\label{eq:spectralVW}
\begin{equation}
	\boldsymbol{\mathcal{V}} : \{ \boldsymbol{e} = \boldsymbol{\nabla} \times \widehat{\boldsymbol{\Psi}}, \quad \boldsymbol{e} \boldsymbol{\cdot} \boldsymbol{1}_n = 0 \ \, \text{on} \ \, \partial V \} \quad \text{and} \quad 
	\boldsymbol{\mathcal{W}}: \{ \boldsymbol{e} = \boldsymbol{\nabla} \widehat{\Phi}, \quad \boldsymbol{e} \boldsymbol{\cdot} \boldsymbol{1}_n = 0 \ \, \text{on} \ \, \partial V \}, 
    \tag{\theequation a,b}
\end{equation}
\end{subequations}
which are not mutually orthogonal with respect to (\ref{eq:scalprod0}).
Nevertheless, spectral decomposition (\ref{eq:helmholtzweighted2}) is often preferable for numerical computations because spaces (\ref{eq:spectralVW}) do not depend on $\rho_0$, contrary to the orthogonal spaces in decomposition (\ref{eq:helmholtzweighted}). 

\subsection{Finite-dimensional approximation}
\label{subsec:modalpb}
The solutions of QEP (\ref{eq:compressibleQEP}) can be accurately obtained as follows. 
We introduce the finite-dimensional form of sub-spaces (\ref{eq:spectralVW}), denoted $\boldsymbol{\mathcal{V}} \, [n\geq1]$ and $\boldsymbol{\mathcal{W}} \, [n\geq2]$.
They are spanned by vectors with components made of Cartesian monomials $x^iy^jz^k$ with $i+j+k\leq n$. 
Their dimensions are \cite{vidal2020acoustics} $\dim \boldsymbol{\mathcal{V}} \, [n\geq1] = n (n+1)(2n+7)/6$ and $\dim \boldsymbol{\mathcal{W}} \, [n\geq2] = n(n+1)(n+2)/6 - 1$.
Then, we seek polynomial expansions of the momentum $\rho_0 \boldsymbol{\zeta}$ in the form
\begin{equation}
	\rho_0 \boldsymbol{\zeta} (\boldsymbol{r}) = \sum_{j=1}^{N} \gamma_{j} \, \boldsymbol{e}_j \, (\boldsymbol{r}), \quad \boldsymbol{e}_j \boldsymbol{\cdot} \boldsymbol{1}_n = 0 \ \, \text{on} \ \, \partial V,
	\label{eq:expandVnWn}
\end{equation}
where $\boldsymbol{\gamma} = (\gamma_{1}, \gamma_{2}, \dots)^T$ is the (complex-valued) state vector and $\{\boldsymbol{e}_j \}$ the (real-valued) polynomial vector elements in $\boldsymbol{\mathcal{V}} \, [n\geq1]$ and $\boldsymbol{\mathcal{W}} \, [n\geq2]$. 
To cover all the admissible polynomial elements, the sum in expansion (\ref{eq:expandVnWn}) goes over indices up to $N = \dim \boldsymbol{\mathcal{V}} \, [n\geq1] + \dim \boldsymbol{\mathcal{W}} \, [n\geq2]$. 
The generation of the basis elements is explained in appendix \ref{appendix:spectral}. 
The displacement $\boldsymbol{\zeta}$ obtained from any polynomial expansions of $\rho_0 \boldsymbol{\zeta}$ diverges when $\rho_0$ vanishes on the boundary (see the electronic supplementary material), so we do not consider the situation $\alpha=1$ in the following.

\begin{figure}
	\centering
	\begin{tabular}{cc}
	\includegraphics[width=0.45\textwidth]{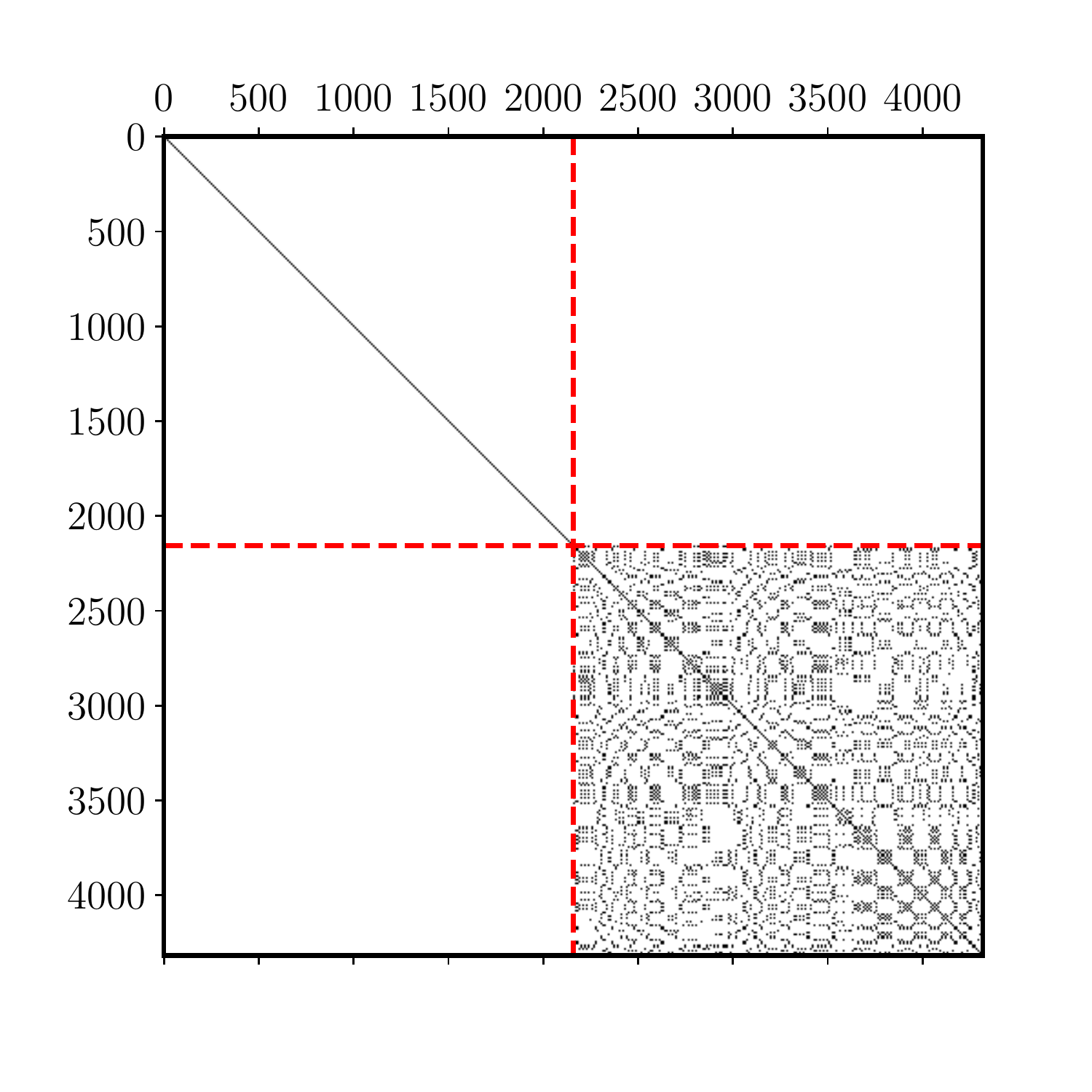} & 
	\includegraphics[width=0.45\textwidth]{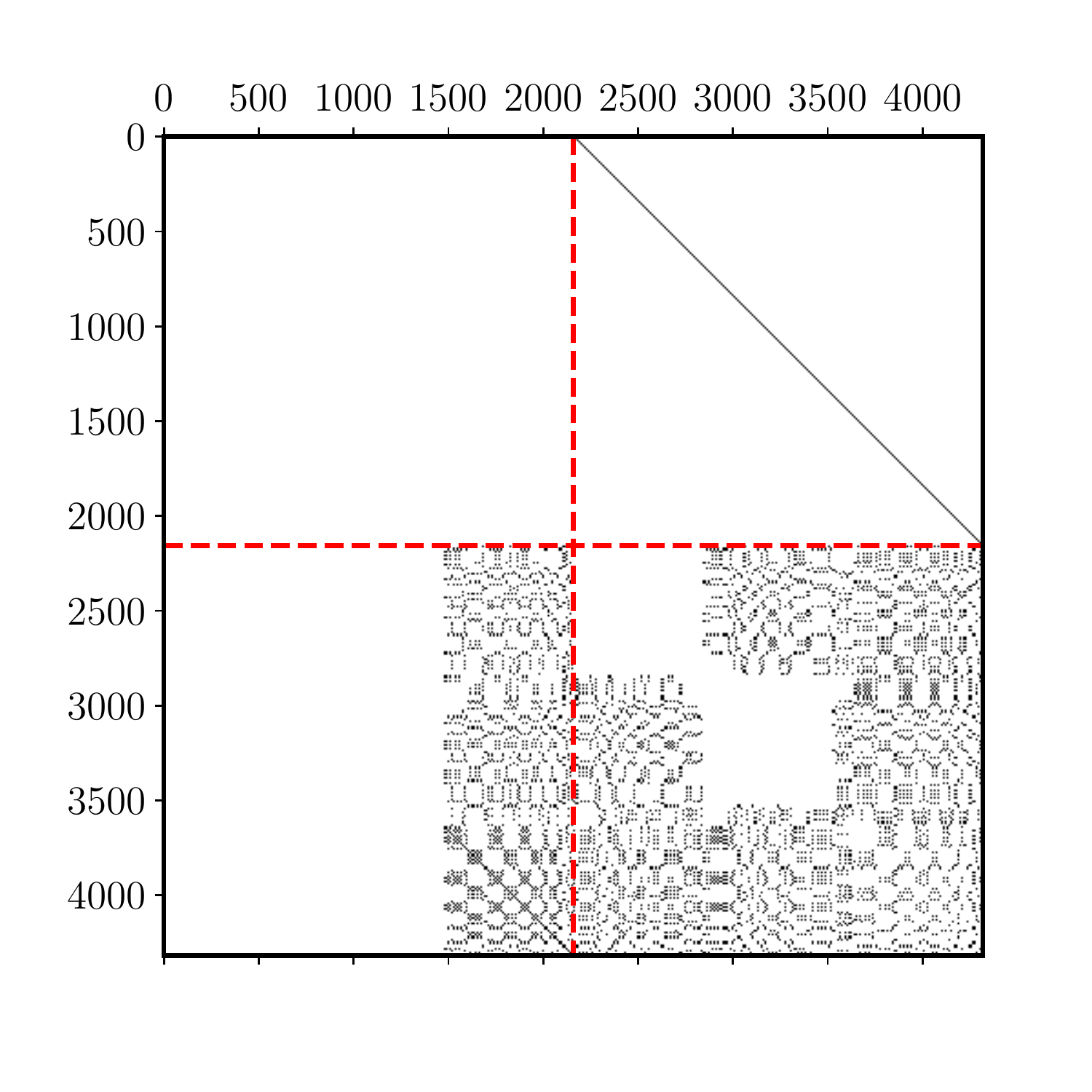} \\ 
	$\boldsymbol{B} \, (= \boldsymbol{B}^\top)$ & $\boldsymbol{A}$ \\
	\end{tabular}
	\caption{Non-zero entries of the sparse matrices $[\boldsymbol{B},\boldsymbol{A}]$ in GEP (\ref{eq:GEP_linearised}). Symbolic computations with $\alpha=0.205$ (Earth-like model) at polynomial degree $n=15$ with the dimension $N = 2159$. Red (dashed) lines show the block structure. (Online version in colour.)}
	\label{fig:MatLRHS}
\end{figure}

We obtain the governing equations for the state vector $\boldsymbol{\gamma}$ by using the method of weighted residuals. 
We substitute truncated expansion (\ref{eq:expandVnWn}) in QEP (\ref{eq:compressibleQEP}) and we project, with respect to inner product (\ref{eq:scalprod0}), the resulting equations onto every basis element $\boldsymbol{e}_i$ to numerically minimize the residual terms (Galerkin method). 
The integral projections give the finite-dimensional QEP 
\begin{equation}
	\left [ \lambda^2 \boldsymbol{M} + \lambda \, \boldsymbol{C}  + \boldsymbol{K} \right ] \boldsymbol{\gamma} = \boldsymbol{0},
	\label{eq:QEPfinite}
\end{equation}
with the eigenvalue-eigenvector pair $[\lambda, \boldsymbol{\gamma}]$. 
The three matrices $[\boldsymbol{M}, \boldsymbol{C}, \boldsymbol{K}]$ in QEP (\ref{eq:QEPfinite}) are real-valued (because the basis elements are real polynomials), and with the non-zero elements
\begin{subequations}
\label{eq:galerkinproj}
\begin{equation}
	M_{ij} = \int_V \boldsymbol{e}_i \boldsymbol{\cdot} \boldsymbol{e}_j \, \mathrm{d}V, \quad 
	C_{ij} = \int_V \boldsymbol{e}_i \boldsymbol{\cdot} \boldsymbol{\mathcal{C}} ( \boldsymbol{e}_j) \, \mathrm{d}V, \quad 
	K_{ij} = \int_V \boldsymbol{e}_i \boldsymbol{\cdot} \rho_0 \,  \boldsymbol{\mathcal{K}} ( \boldsymbol{e}_j) \, \mathrm{d}V.
    \tag{\theequation a--c}
\end{equation}
\end{subequations}
The matrix $\boldsymbol{M}$ is positive definite, and $\boldsymbol{C}$ is skew-Hermitian. 
Moreover, since expressions (\ref{eq:galerkinproj}) only involve Cartesian monomials, they can be evaluated analytically \cite[see formula (50)]{lebovitz1989stability}. 

We have modified the numerical code \textsc{shine}, initiated in Vidal \emph{et al.} \cite{vidal2020acoustics}, to implement the aforementioned spectral algorithm. 
To reduce the condition number of the matrices, which affects the numerical accuracy, we normalise the basis elements such that $M_{ii} = 1$ in (\ref{eq:galerkinproj}a). 
Then, we convert QEP (\ref{eq:QEPfinite}) into the generalized eigenvalue problem (GEP) of double size $2N$
\begin{equation}
	\lambda \underbrace{\begin{pmatrix}
		\boldsymbol{I} & 0 \\
		0 & \boldsymbol{M} \\
	\end{pmatrix}}_{\boldsymbol{B}} \,
	\begin{pmatrix}
		\boldsymbol{\gamma} \\
		\lambda \boldsymbol{\gamma}  \\
	\end{pmatrix}
	= \underbrace{\begin{pmatrix}
		0 & \boldsymbol{I} \\
		-\boldsymbol{K} & -\boldsymbol{C} \\
	\end{pmatrix}}_{\boldsymbol{A}} \,
	\begin{pmatrix}
		\boldsymbol{\gamma} \\
		\lambda \boldsymbol{\gamma} \\
	\end{pmatrix},
	\label{eq:GEP_linearised}
\end{equation}
with $\boldsymbol{I}$ the identity matrix and $[\boldsymbol{A},\boldsymbol{B}]$ two matrices. 
GEP (\ref{eq:GEP_linearised}) has formally $2N$ eigenvalues (possibly degenerate), although we have $N$ unknowns in expansion (\ref{eq:expandVnWn}). 
Yet, by analogy with the infinite-dimensional configuration, there are only $N$ distinct solutions because the real-valued eigenfrequencies come in pairs $[\omega, -\omega]$ \cite{tisseur2001quadratic}. 
The matrix structure of GEP (\ref{eq:GEP_linearised}) is shown in figure \ref{fig:MatLRHS}.
The matrix $\boldsymbol{B}$ is positive-definite but not diagonal, since the basis elements are not orthogonal. 
Finally, we truncate the polynomial expansion at $n\leq 20$ and use double-precision arithmetic. 

\section{Numerical results}
\label{sec:numerics}
We can now investigate the numerical properties of the normal modes in isentropic interiors. 
Based on the discussion in \S\ref{sec:model}\ref{subsec:planets}, where we have estimated the amplitude of the various effects, we only consider isentropic profiles for spheroidal geometries (i.e. $a=b \geq c$), and explore the planetary values $M_\Omega \ll 1$ and $0 \leq \alpha < 1$. 
We have two families of modes in the rotating regime \cite{valette1989spectre}. 
The highest-frequency family represents the acoustic modes, which are known to be weakly sensitive to global rotation (see below). 
The other family is made of the inertial modes (when $M_\Omega \leq 10^{-1}$), which are sustained by the Coriolis force and belong to the frequency range $|\omega| < 2 M_\Omega$. 
We describe the main properties of the different modes in the next subsections. 

\subsection{Acoustic modes}
\label{subsec:sound}
The acoustic modes correspond to the discrete part of the spectrum \cite{valette1989spectre}, which is made up of all the eigenvalues of finite multiplicity that are isolated in the spectrum. 
This guarantees that, in considering high enough polynomial degrees $n$, there are no spurious modes (any polynomial solutions will converge towards the proper acoustic modes). 
We set $n=20$ in the following, to get an excellent convergence of the acoustic modes of interest. 

\subsubsection{Non-rotating regime}
\begin{figure}
    \centering
    \begin{tabular}{cc}
    \includegraphics[width=0.49\textwidth]{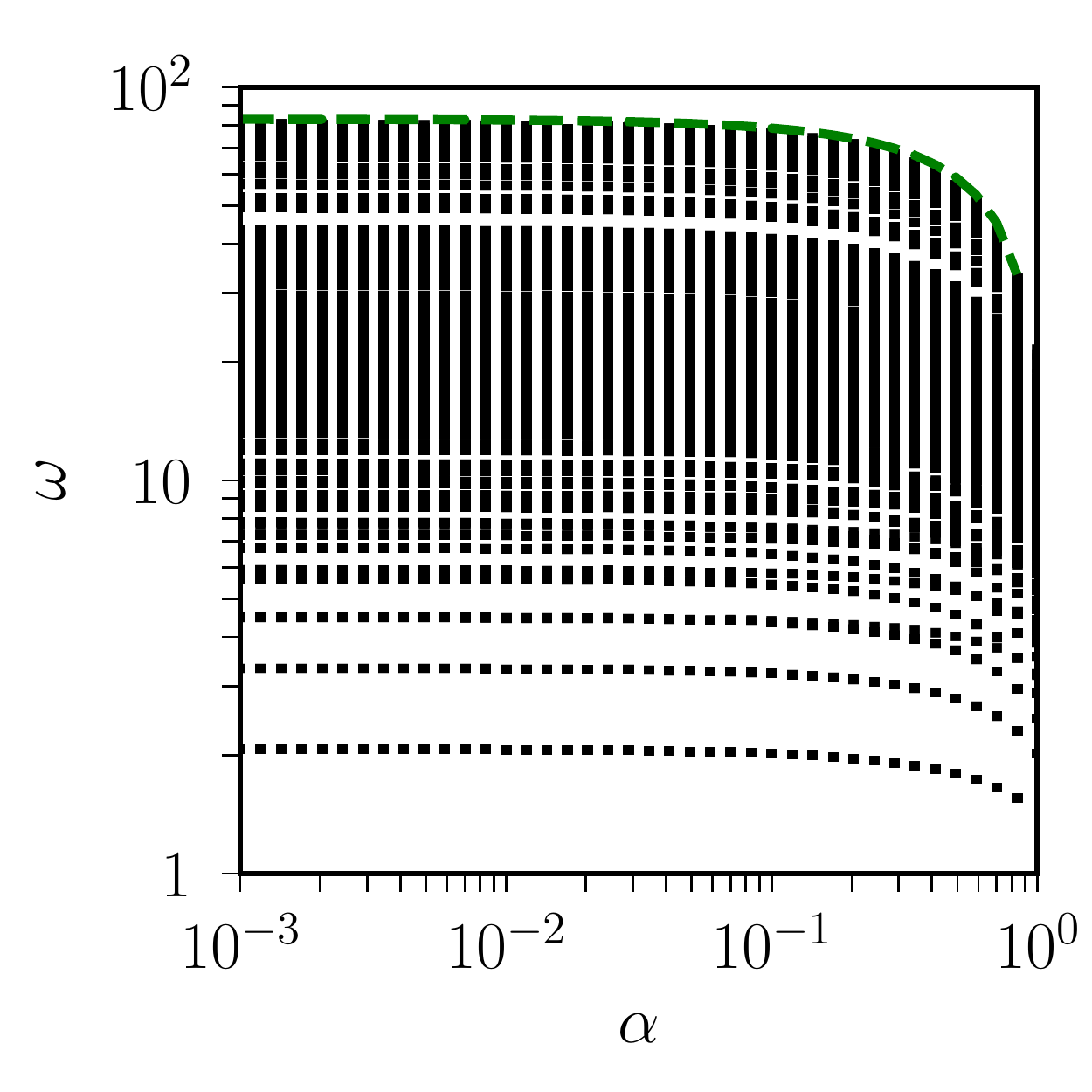} &
    \includegraphics[width=0.49\textwidth]{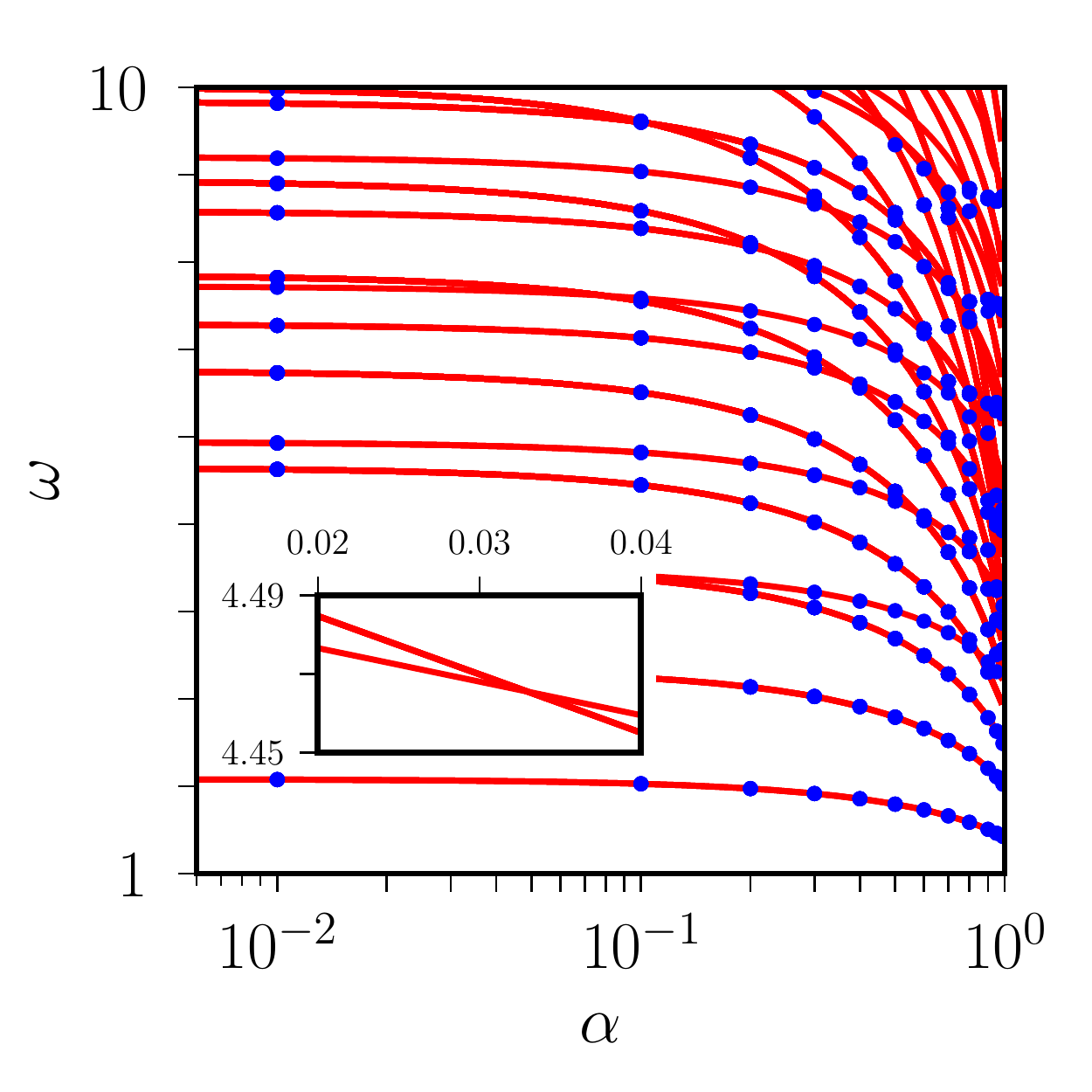} \\
    \end{tabular}
    \caption{Angular frequency $\omega$ of non-rotating acoustic modes, as a function of $\alpha$ in spheres with $n=20$.  (a) Black points: polynomial solutions. Green dashed line: asymptotic scaling (\ref{eq:scalingacoustics}), where the prefactor has been fixed to match the highest-frequency acoustic branch. (b) Red curves: polynomial solutions in the frequency range $1 \leq \omega \leq 10$. Blue points: finite-element solutions of equation (\ref{eq:eqacoustics}). Inset shows accidental degeneracy for $3\times 10^{-2} \leq \alpha \leq 4\times 10^{-2}$. (Online version in colour.)}
    \label{fig:nonrotsound}
\end{figure}

We start with the non-rotating regime.  
The shorter is the wavelength of an acoustic mode, the higher is its frequency, such that the latter is generally well approximated by short-wavelength theory $|\omega| \simeq C_0 \, ||\boldsymbol{k}||$ for the highest-frequency modes (with $\boldsymbol{k}$ the local wave vector).
Given profile (\ref{eq:backgroundC0}) for the isentropic speed of sound, the highest-frequency acoustic modes should thus obey the asymptotic scaling
\begin{equation}
    \omega \propto \sqrt{1-\alpha},
    \label{eq:scalingacoustics}
\end{equation}
which indicates that the frequency of the acoustic modes should decrease with compressibility.
We show in figure \ref{fig:nonrotsound} the spectrum of the non-rotating modes in the sphere, and find indeed a very good agreement with scaling (\ref{eq:scalingacoustics}) in the high-frequency regime (figure \ref{fig:nonrotsound}a). 

To get a more quantitative benchmark, we compute the non-rotating acoustic modes with another numerical method. 
We recast (\ref{eq:compressible2}) as 
\cite[see equation (4)]{chaljub2004spectral}
\begin{equation}
    \lambda \left ( \lambda \boldsymbol{\zeta} + 2 M_\Omega \, \boldsymbol{1}_z \times \boldsymbol{\zeta} \right ) = \boldsymbol{\nabla} \left [ C_0^2 \, \boldsymbol{\nabla} \boldsymbol{\cdot} \boldsymbol{\zeta}  + \boldsymbol{\zeta} \boldsymbol{\cdot} \boldsymbol{g} \right ] + C_0^2 \, (\boldsymbol{\nabla} \boldsymbol{\cdot} \boldsymbol{\zeta}) \, \boldsymbol{S}_0,
    \label{eq:compressiblechaljub}
\end{equation}
with the (vectorial) Schwarzschild discriminant $\boldsymbol{S}_0 = (1/\rho_0) \, {\boldsymbol{\nabla} \rho_0} - {\boldsymbol{g}}/{C_0^2}$ \cite{dyson1979perturbations}. 
As explained in appendix \ref{appendix:spectral}, the non-rotating acoustic modes are exactly described in isentropic interiors (with $\boldsymbol{S}_0=\boldsymbol{0}$) by the potential term in decomposition (\ref{eq:helmholtzweighted}). 
We thus seek $\boldsymbol{\zeta} = \boldsymbol{\nabla} \Phi$, and equation (\ref{eq:compressiblechaljub}) reduces to the scalar Helmholtz-like equation
\begin{equation}
    \lambda^2 \Phi = C_0^2 \, \nabla^2 \Phi + (\boldsymbol{\nabla} \Phi) \boldsymbol{\cdot} \boldsymbol{g}, \quad \boldsymbol{\nabla} \Phi \boldsymbol{\cdot} \boldsymbol{1}_n = 0 \ \, \text{on} \ \, \partial V.
    \label{eq:eqacoustics}
\end{equation}
We solve equation (\ref{eq:eqacoustics}) with standard finite-element computations, performed with the commercial software COMSOL.
We use the built-in acoustic solver, in which the additional term $(\boldsymbol{\nabla} \Phi) \boldsymbol{\cdot} \boldsymbol{g}$ is modelled through a source term.
The ellipsoidal domain is discretised with an unstructured mesh made of tetrahedral Lagrange elements. 
We have used cubic Lagrange elements (P3) for $\Phi$, with a total number of $242 \, 865$ degrees of freedom.
We superimpose the COMSOL computations onto the polynomial solutions of the full QEP in figure \ref{fig:nonrotsound}b.  
We obtain an excellent quantitative agreement, even if the governing equations and numerical methods are very different. 
This successfully validates the accuracy of the polynomial description. 

\begin{figure}
    \centering
    \begin{tabular}{cc}
    \includegraphics[width=0.49\textwidth]{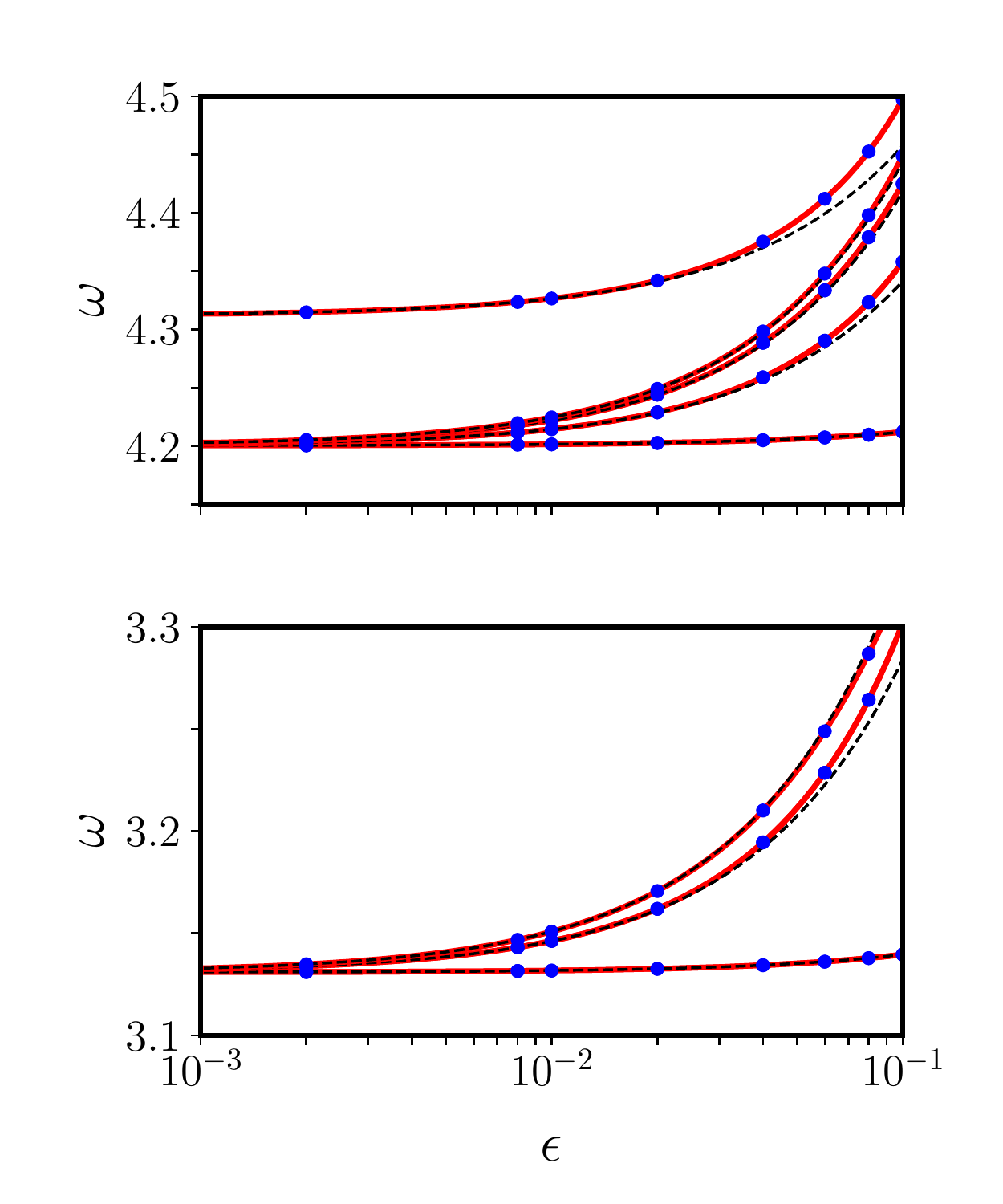} &
    \includegraphics[width=0.49\textwidth]{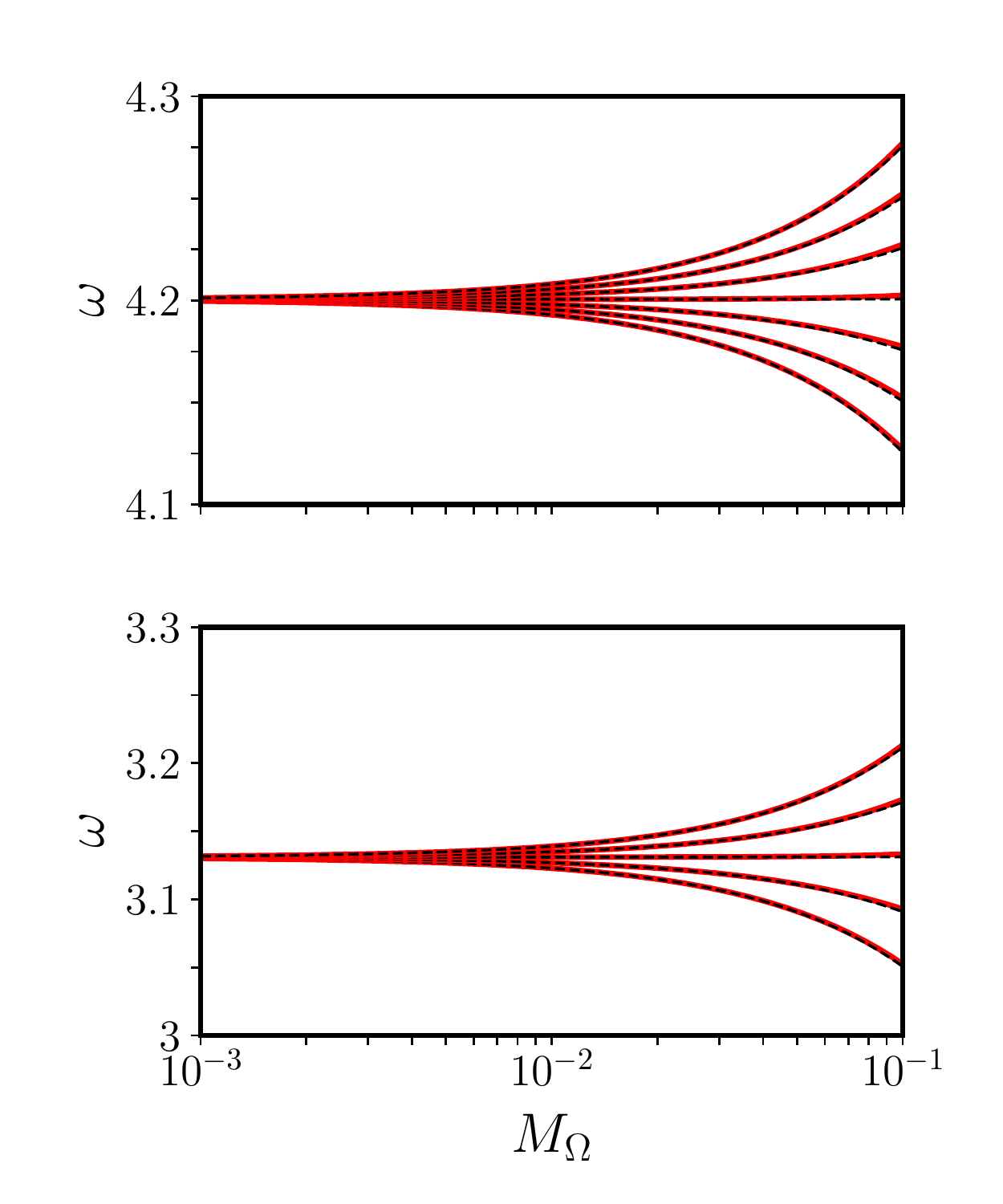} \\
    \end{tabular}
    \caption{Angular frequency, as a function of (a) $\epsilon=1-c/a$ in non-rotating spheroids and (b) $M_\Omega$ in spheres, for $\alpha=0.205$.
    Solid (red) curves: polynomial solutions with $n=20$. Dashed (black) curves: linear fits estimated from polynomial values in the range $\epsilon \leq 10^{-2}$ (a) and $M_\Omega \leq 10^{-2}$ (b), as predicted by first-order perturbation theories for the non-rotating sphere.
    Acoustic modes with the spherical harmonic degree $l=2$ (bottom panel), and $l=3$ (top panel). Blue points: finite-element solutions of equation (\ref{eq:eqacoustics}). (Online version in colour.)}
    \label{fig:sound2}
\end{figure}

In non-rotating and uniform-density spheres, the acoustic modes with different azimuthal wave numbers $m$, but with the same latitudinal structures, have the same angular frequency (degeneracy).
This is a consequence of the assumed spherical symmetry of the background state, which is independent of the chosen polar axis.
We do obtain these degeneracies in figure \ref{fig:nonrotsound}, where the corresponding modes are indeed superimposed. 
Moreover, we uncover new occurrences of accidental degeneracy for modes with different spatial structures when $\alpha\neq0$. 
Degeneracy is clearly observed for high-frequency modes when $\alpha\geq 10^{-1}$, so firmly in the planetary range of values (table \ref{table:planets}), but some lower-frequency modes are also affected (see inset in figure \ref{fig:nonrotsound}b, for $3 \times 10^{-2} \leq \alpha \leq 4 \times 10^{-2}$).

Next, we quantify in figure \ref{fig:sound2}a the flattening effects for a few large-scale acoustic modes. 
It is known that a first-order perturbation treatment is generally inaccurate to predict the angular frequency of the acoustic modes in strongly flattened bodies \cite{reese2006acoustic,vidal2020acoustics}.
Indeed, second-order corrections in $\epsilon^2$ can become non-negligible when $\epsilon \gtrsim 10^{-1}$ (e.g. for the upper branch in the top panel), which are realistic values for rapidly rotating planets (table \ref{table:planets}) or experiments \cite{su2020zoro}.  

\subsubsection{Coriolis effects}
\begin{figure}
    \centering
    \begin{tabular}{cc}
    \includegraphics[width=0.49\textwidth]{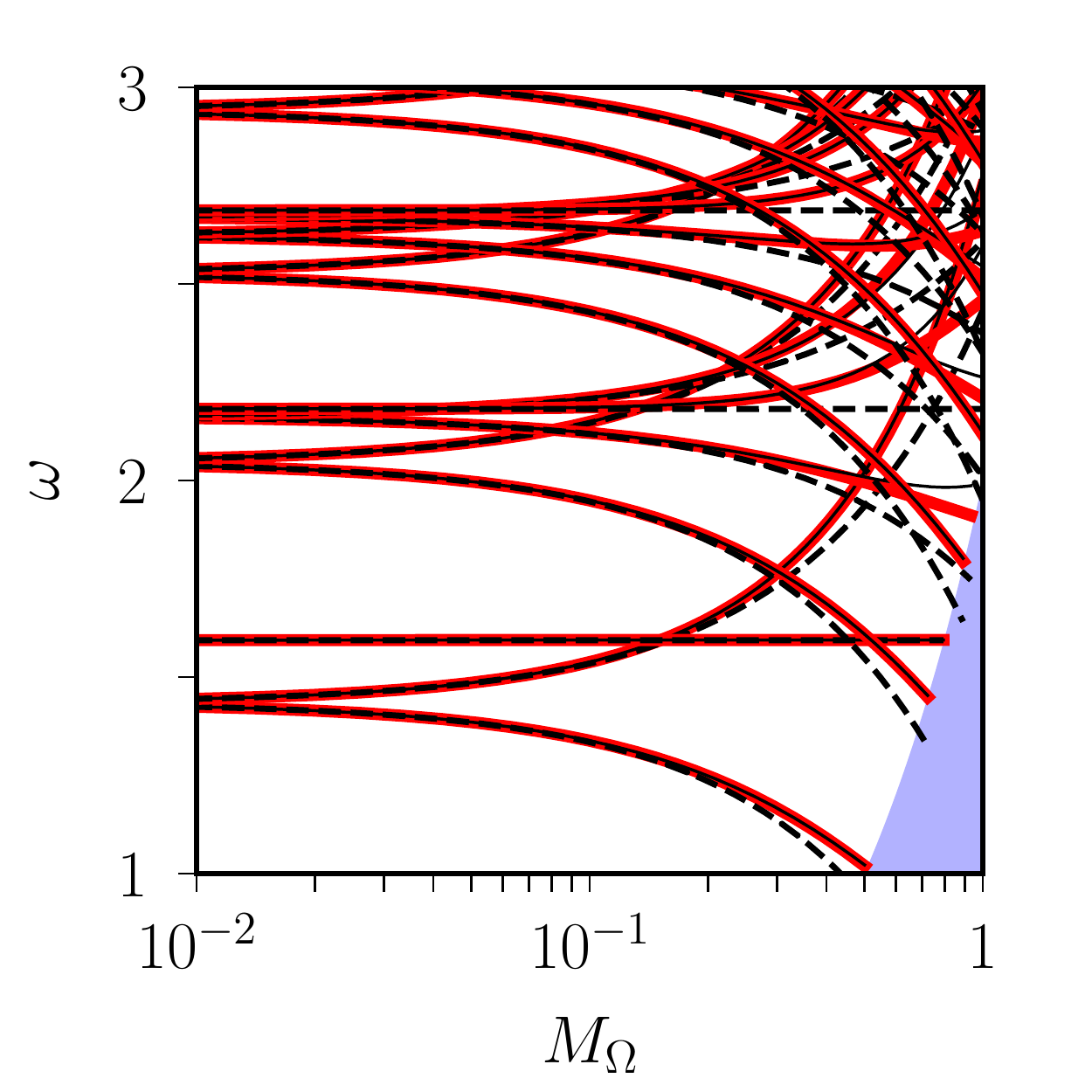} &
    \includegraphics[width=0.4\textwidth]{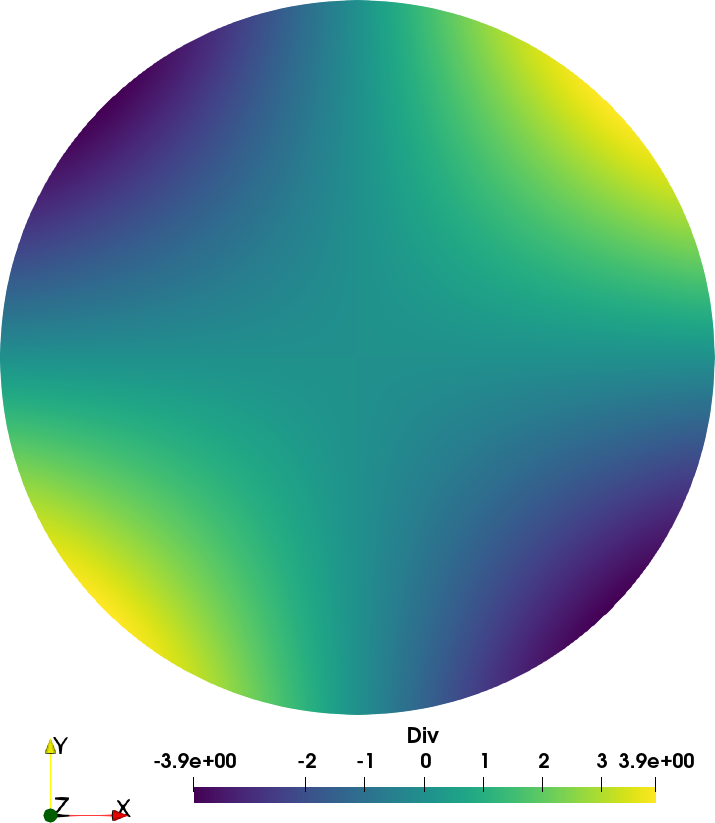} \\
    (a) & (b) \\
    \end{tabular}
    \caption{(a) Angular frequency, as a function $M_\Omega$, in the spheroid with $\epsilon=0.1$ and $\alpha=0.98$. Solid (red) curves: polynomial solutions with $n=20$. Black curves: linear (dashed) and quadratic (thin, solid) predictions, estimated from polynomial values in the range $10^{-3} \leq M_\Omega \leq 10^{-2}$, as given by perturbation theory for the non-rotating spheroid. 
    The inertial modes in the range $|\omega| < 2 M_\Omega$ (coloured region) have been removed for clarity.
    (b) Equatorial slice of density perturbation $\boldsymbol{\nabla} \boldsymbol{\cdot} (\rho_0 \boldsymbol{\zeta})$ for the mode $\omega \simeq 1.949$ in the rotating spheroid with $M_\Omega=10^{-1}, \epsilon=0.1$ and $\alpha=0.98$.}
    \label{fig:sound3}
\end{figure}

We reintroduce global rotation by setting $M_\Omega \neq 0$. 
The acoustic modes can no longer be sought with equation (\ref{eq:eqacoustics}), such that only QEP (\ref{eq:compressibleQEP}) can be solved. 
The acoustic spectrum is illustrated in figure \ref{fig:sound2}b with $\alpha=0.205$. 
Similar behaviours are found for other values of $\alpha$, and so the results are omitted for the sake of concision.
The Coriolis force has measurable effects on the acoustic modes when $M_\Omega \geq 10^{-2}$, lifting the degeneracy of the non-rotating spherical modes. 
The non-rotating (degenerate) modes, characterized by the spherical harmonic degree $l$, split into $2l+1$ modes as observed here. 
However, the short-wavelength acoustic modes are less affected by the Coriolis force than the large-scale modes \cite{reese2006acoustic}.
Thus, the Coriolis splitting is generally smaller in amplitude for the high-frequency acoustic modes than for the low-frequency ones (not shown). 

The Coriolis splitting is often modelled by first-order perturbation theory \cite{backus1961rotational}.
We have superimposed in figure \ref{fig:sound2}b the linear fits in $M_\Omega$, which would be given by such a theory. 
Actually, the linear predictions reproduce fairly the (non-perturbative) polynomial computations in the range $M_\Omega \ll 10^{-1}$. 
However, rapidly rotating gas giants are characterized by much larger values of $M_\Omega$ (table \ref{table:planets}). 
This situation also occurs in the experimental context, where larger values $M_\Omega \sim 10^{-1}$ can be achieved \cite{su2020zoro}. 
Moreover, global rotation can also strongly flatten the fluid boundary due to centrifugal effects in the planetary context \cite{zhang2017shape}, but treating separately ellipticity and Coriolis effects has not proven accurate enough (e.g. for experimental conditions \cite{vidal2020acoustics,su2020zoro}). 
Therefore, ellipticity and Coriolis effects should be modelled together. 
We show in figure \ref{fig:sound3}a the results in the rotating spheroid, considering a Jupiter-like model with $\epsilon=0.1$ and $\alpha=0.98$. 
We have superimposed the linear predictions $\omega \simeq a_0 + a_1 M_\Omega$ (dashed curves) for the non-rotating spheroid. 
The linear estimates strongly depart from the non-perturbative polynomial solutions for the lowest-frequency acoustic modes in the range $M_\Omega \gtrsim 10^{-1}$.
This clearly shows that first-order perturbation theory is inaccurate to model the Coriolis effects for rapidly rotating Jupiter-like models, but a second-order theory $\omega \simeq a_0 + a_1 M_\Omega + a_2 M_\Omega^2$ for the non-rotating spheroid (thin solid lines in figure \ref{fig:sound3}a) appears sufficient for all the acoustic modes in the planetary range.

Finally, we illustrate in figure \ref{fig:sound3}b the spatial structure of an acoustic mode in the rotating spheroid ($M_\Omega=10^{-1}, \epsilon=0.1$) with $\alpha=0.98$. 
The acoustic modes can keep large-scale structures in the bulk (at least for the lowest-frequency ones), even in strongly compressible interiors. 

\subsection{Inertial modes}
\label{subsec:inertial}
\begin{figure}
    \centering
    \begin{tabular}{cc}
    \includegraphics[width=0.49\textwidth]{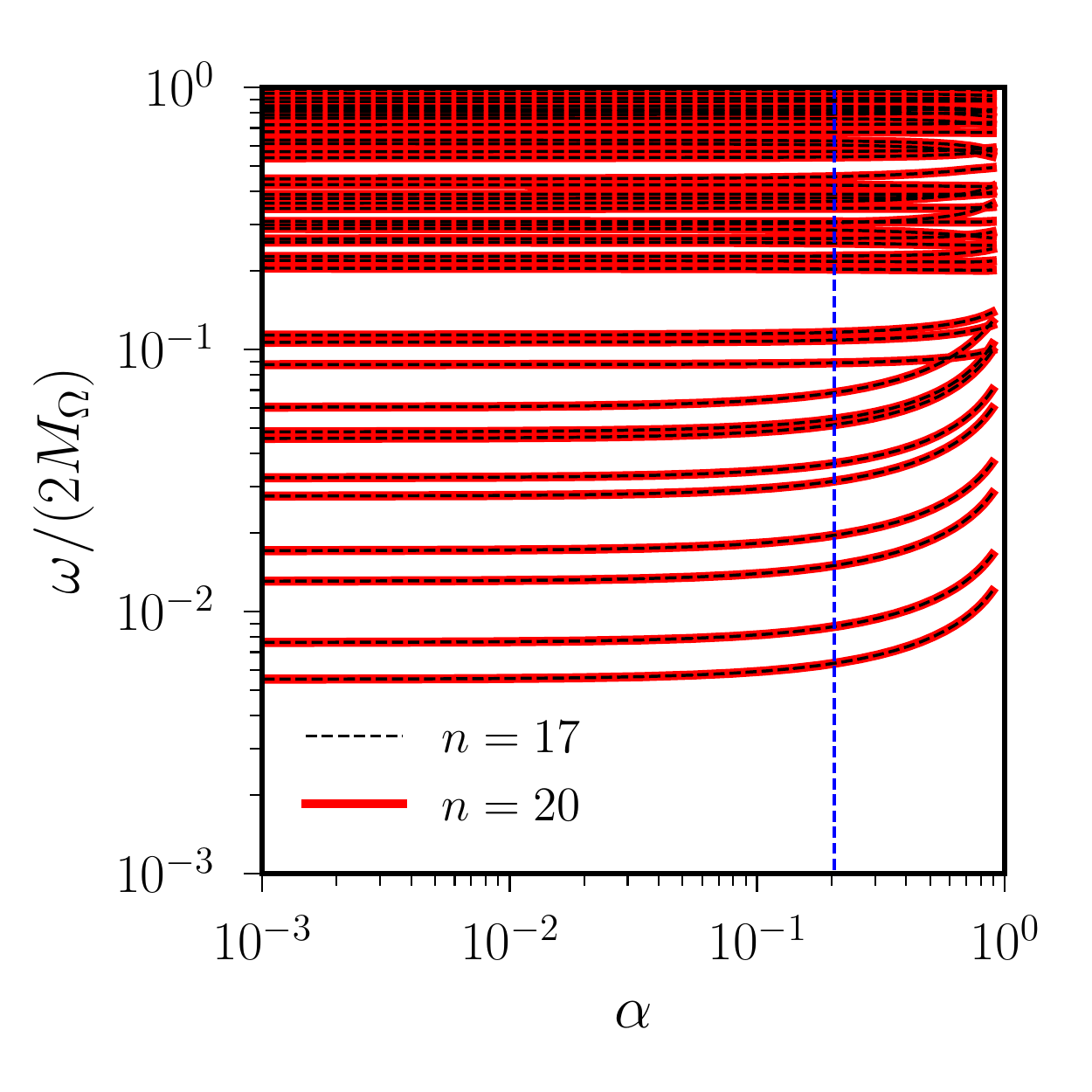} &
    \includegraphics[width=0.49\textwidth]{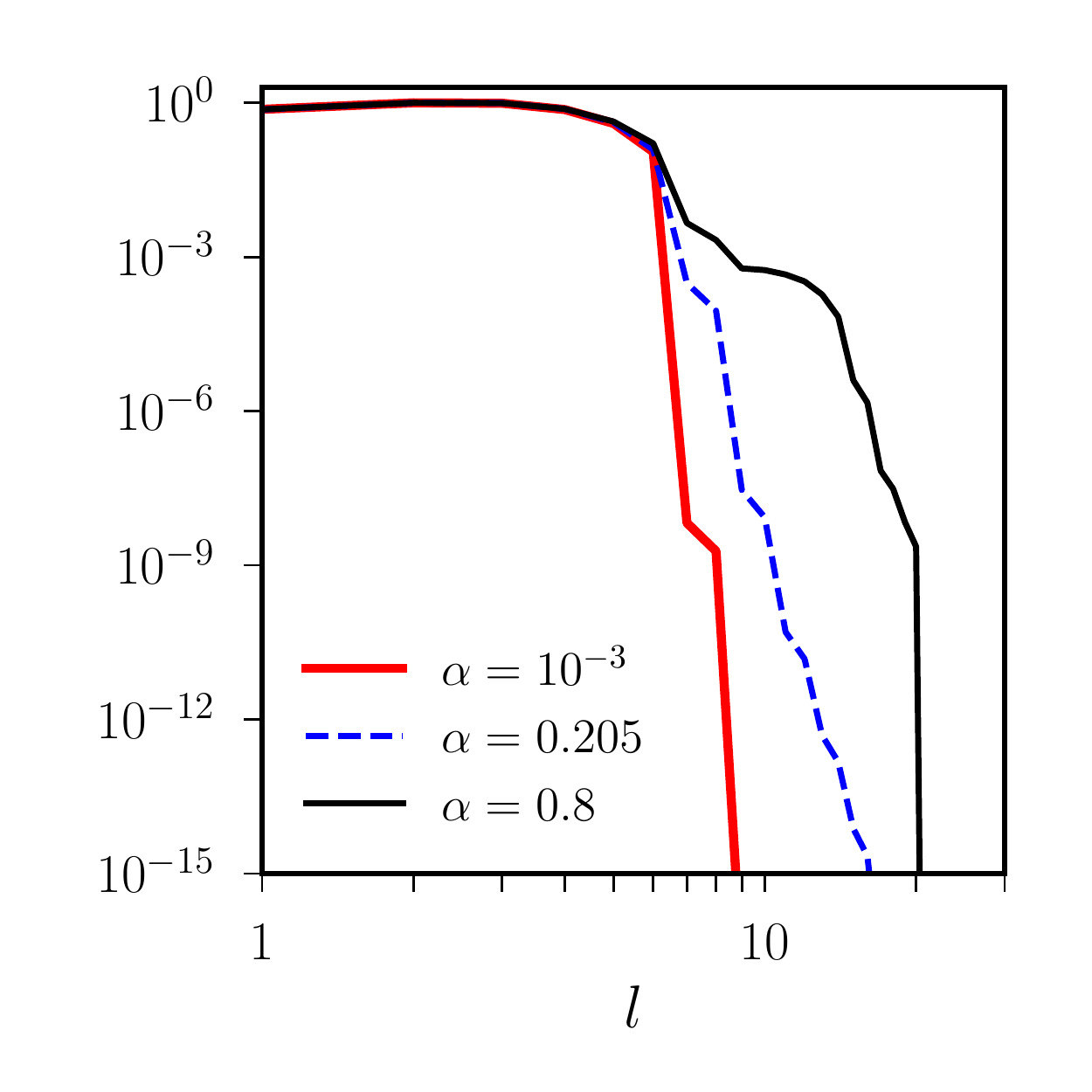} \\
    (a) & (b) \\
    \end{tabular}
    \caption{Convergence of inertial modes in the sphere with $M_\Omega=10^{-2}$. (a) Normalized angular frequency $\omega/(2M_\Omega)$, as a function of $\alpha<1$, for a subset of modes. Vertical dashed line shows the Earth-like value $\alpha=0.205$. (b) Normalized power spectrum (radial average) of $\boldsymbol{\zeta}$, as a function of the spherical harmonic degree $l$, for the non-toroidal inertial mode shown in figure \ref{fig:IM2} with $n=20$. (Online version in colour.) 
    }
    \label{fig:IM0}
\end{figure}

We now focus on the inertial modes, which belong to the frequency range $|\omega|<2M_\Omega$ when $M_\Omega \ll 1$ \cite{valette1989spectre}. 
The inertial modes are only known to be exact polynomials in uniform-density fluid-filled ellipsoids \cite{backus2017completeness,ivers2017enumeration}. 
Hence, we have only considered compressible solutions that clearly exhibit numerical convergence, as shown for instance by the absence of any significant changes on the frequency between the degrees $n=17$ and $n=20$ in figure \ref{fig:IM0}a. 
Their spatial spectra are also well converged (figure \ref{fig:IM0}b). 
We first consider the toroidal modes (or r-modes), and then we survey the other (non-toroidal) inertial modes.  

\subsubsection{Toroidal modes}
The r-modes are a subset of inertial modes, made of purely toroidal motions in the incompressible theory (i.e. they do not exhibit any radial motions in the rotating sphere, and similarly in the ellipsoid). 
Their angular frequency, in rotating incompressible (and anelastic) spheres, is given in compressible units by \cite{wu2005origin}
\begin{equation}
    \frac{|\omega_{0}|}{M_\Omega} = \frac{2}{m+1},
    \label{eq:wrossby}
\end{equation}
with the azimuthal number $m\geq1$ measured from the rotation axis. 
The r-modes are not entirely toroidal in compressible models, such that formula (\ref{eq:wrossby}) is no longer exact when $\alpha \neq 0$ but remains the leading-order term \cite{papaloizou1978non}. 

\begin{figure}
    \centering
    \begin{tabular}{cc}
    \includegraphics[width=0.49\textwidth]{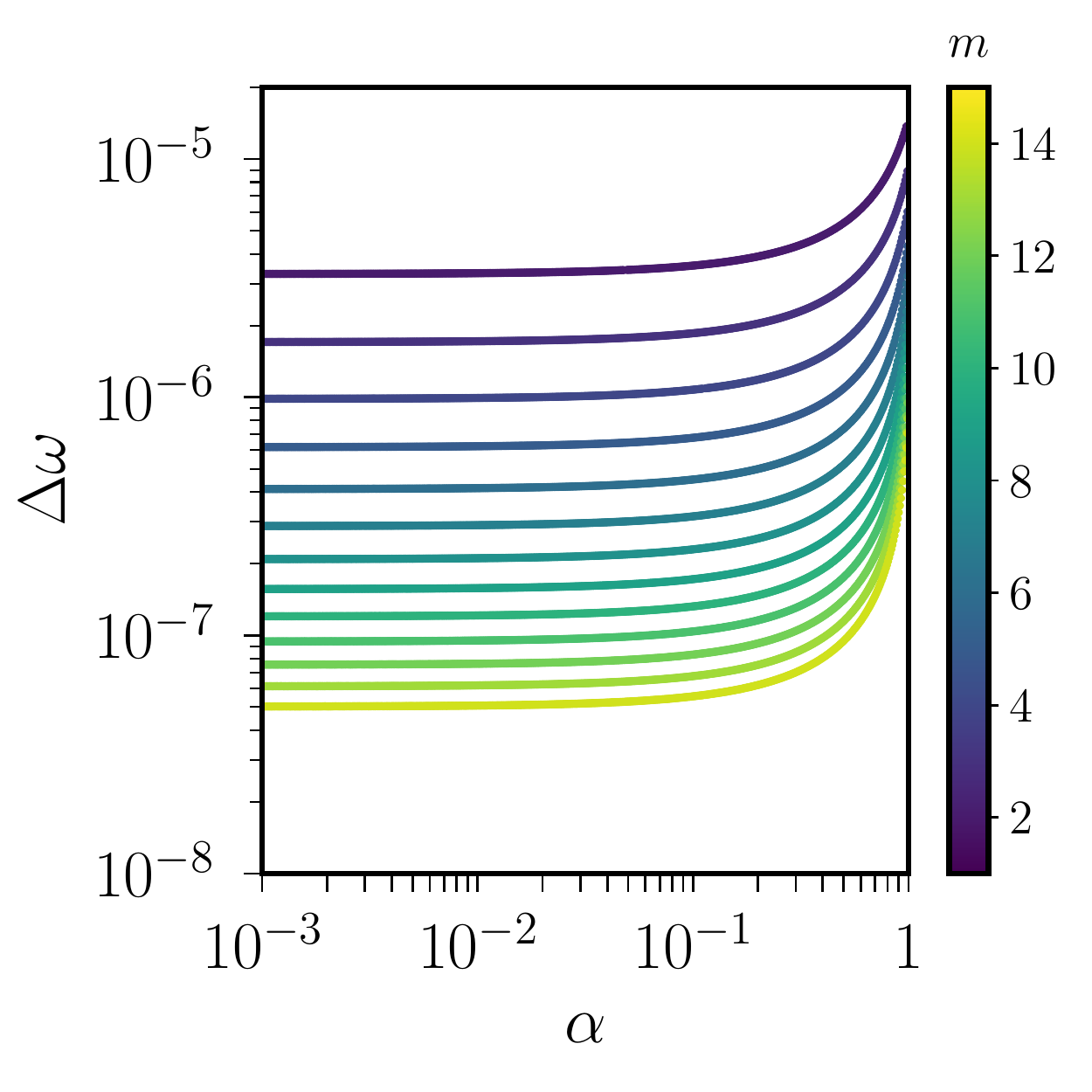} &
    \includegraphics[width=0.49\textwidth]{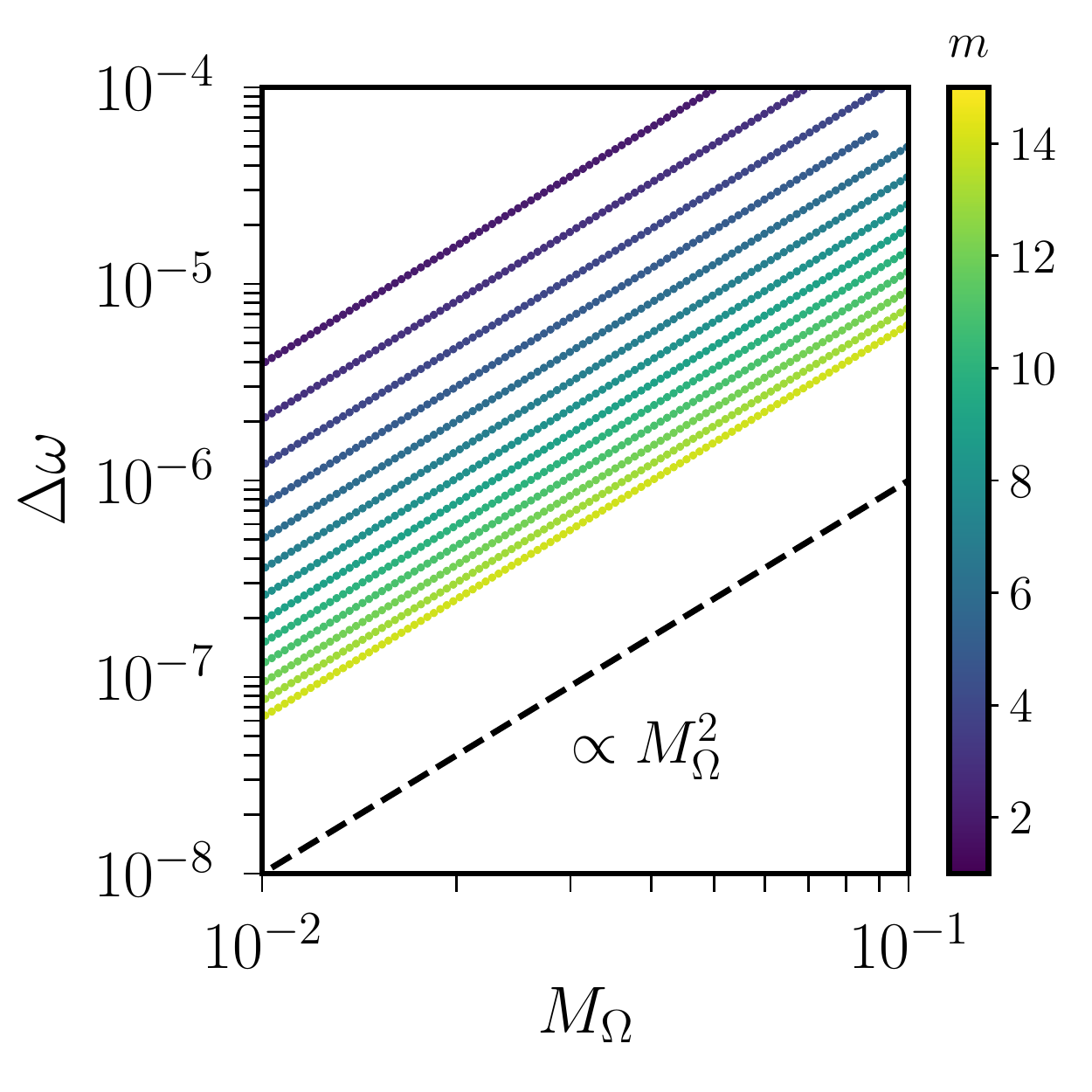} \\
    (a) & (b) \\
    \end{tabular}
    \caption{Variations $\Delta \omega$ for the r-modes in rotating compressible spheres, as a function of $\alpha<1$ with $M_\Omega=10^{-2}$ in (a) and of $M_\Omega$ with $\alpha=0.205$ in (b). Polynomial solutions $n=20$.   
    Colour bar shows the azimuthal number $2\leq m \leq15$. (Online version in colour.)
    }
    \label{fig:rossby1}
\end{figure}

The quantity $\Delta \omega = |\omega - \omega_{0}|/|\omega_{0}|$, measuring the departure from the incompressible solutions $\omega_0$, is shown in figure \ref{fig:rossby1} for some spherical r-modes.  
Their frequency is only weakly modified by compressibility (figure \ref{fig:rossby1}a), with $\Delta \omega \leq 10^{-5}$ at $M_\Omega=10^{-2}$, and the variations are smaller when $m$ is increased. 
The frequency is also slightly modified in compressible interiors when $M_\Omega$ is increased, in agreement with the expected scaling $\Delta \omega \propto M_\Omega^2$ at next order \cite[see equation (3.14b) in the inertial frame]{papaloizou1978non}. 
The r-modes also survive in the coreless triaxial ellipsoid with our barotropic reference state (not shown), because the isopycnics are self-similar ellipsoidal shells that coincide with the boundary. 
Hence, we directly infer from the incompressible theory \cite{vantieghem2014inertial} that planetary ellipticity values $\epsilon \ll 1$ are only responsible for small frequency variations (not shown). 

Finally, we illustrate in figure \ref{fig:IM1}a the spatial structure of the r-modes in the sphere (with $M_\Omega=10^{-2}$). 
We have shown a $m=3$ mode, but similar results are found for the other r-modes.
The r-modes are nearly anelastic (i.e. with $\boldsymbol{\nabla} \boldsymbol{\cdot} (\rho_0 \boldsymbol{\zeta}) \simeq 0$, not shown), and so do not produce any significant density perturbations. 
Their structure is barely modified by compressibility, since a nearly similar structure is found when $\alpha=0$ (not shown). 

\subsubsection{Non-toroidal modes}
\begin{figure}
    \centering
    \begin{tabular}{cc}
    \includegraphics[trim={0.5cm 4.5cm 2cm 3cm},clip,width=0.41\textwidth]{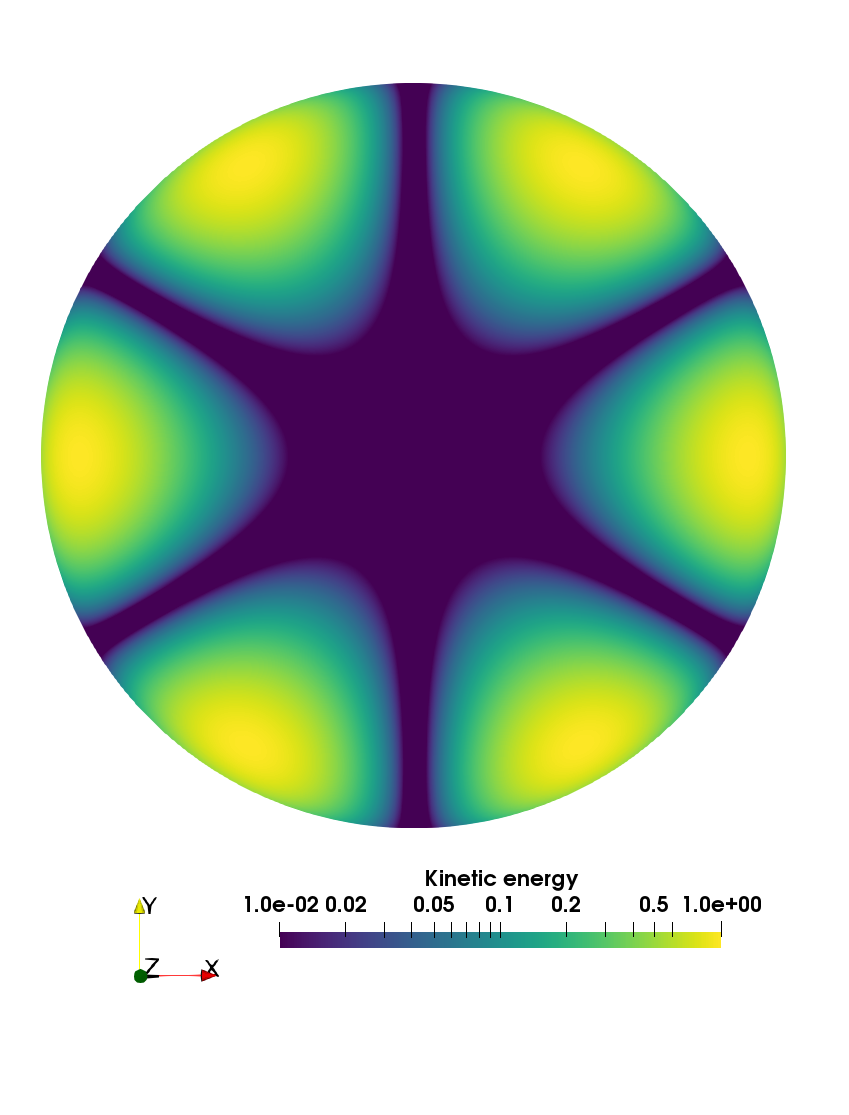} &
    \includegraphics[width=0.49\textwidth]{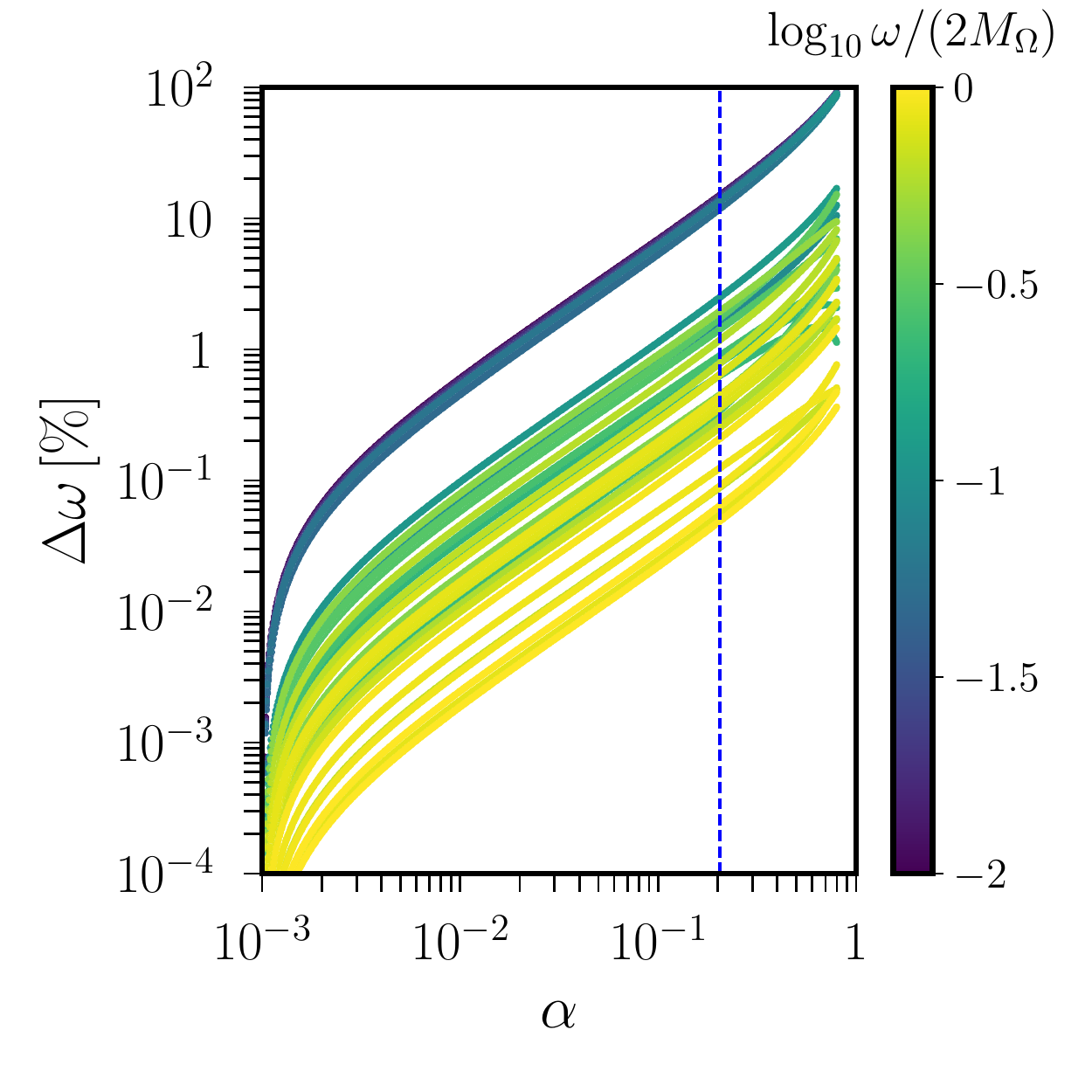} \\
    (a) & (b) \\
    \end{tabular}
    \caption{(a) Kinetic energy (equatorial slice) of spherical r-mode $m=3$ at $n=20$ with $M_\Omega = 10^{-2}$ and $\alpha=0.93$. Colour bar is normalized such that $\max \rho_0 ||\boldsymbol{\zeta}||^2 = 1$. 
    (b) Departure $\Delta \omega$ of angular frequency from incompressible value $\omega_0$ (estimated at $\alpha=10^{-3}$) for a subset of non-toroidal spherical inertial modes with $n=20$ (figure \ref{fig:IM0}a), as a function of $\alpha$ with $M_\Omega=10^{-2}$. 
    Colour bar shows ratio $\omega/(2M_\Omega)$ in logarithmic scale. Vertical dashed line shows the Earth-like value $\alpha=0.205$. 
    (Online version in colour.) 
    }
    \label{fig:IM1}
\end{figure}

\begin{figure}
    \centering
    \begin{tabular}{cc}
    \includegraphics[trim={2cm 4cm 2cm 4cm},clip,width=0.46\textwidth]{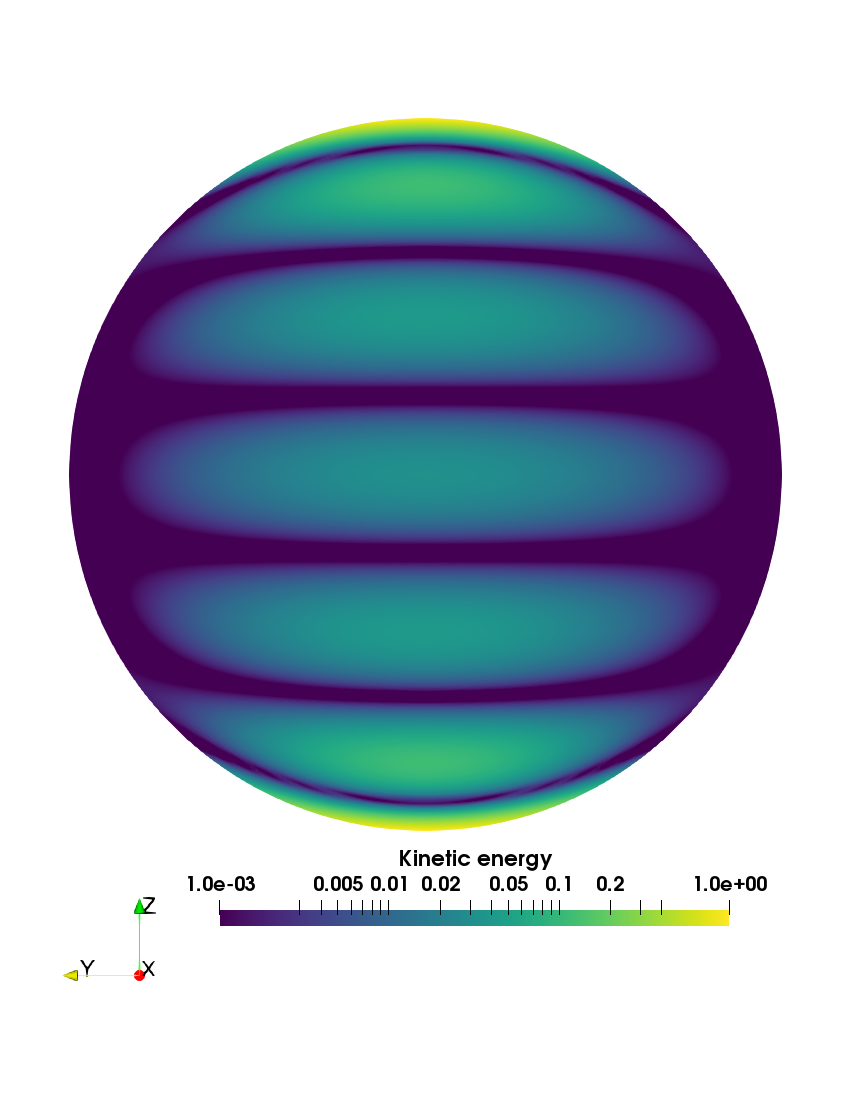} &
    \includegraphics[trim={2cm 4cm 2cm 4cm},clip,width=0.46\textwidth]{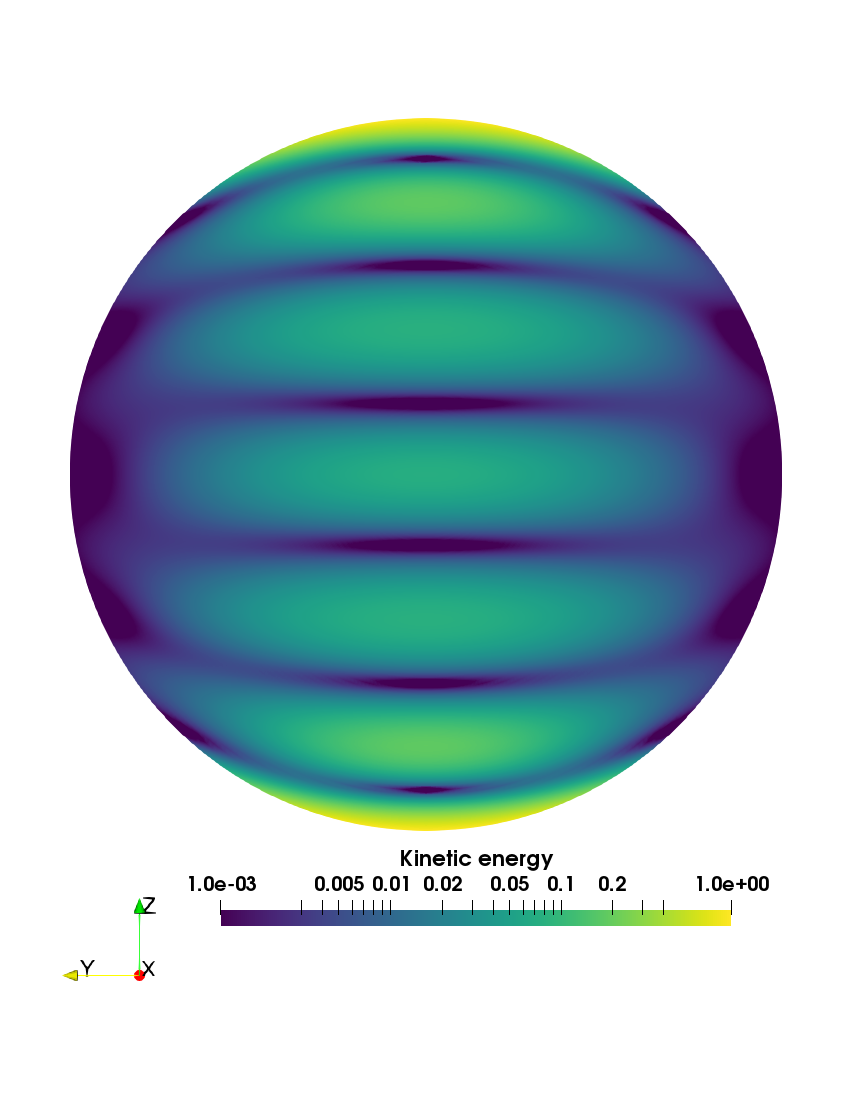} \\
    (a) & (b) \\
    \end{tabular}
    \caption{Kinetic energy $\rho_0 ||\boldsymbol{\zeta}||^2$ of an equatorially symmetric non-toroidal inertial mode in the sphere with $M=10^{-2}$. Amplitude is normalised such that $\max \rho_0 ||\boldsymbol{\zeta}||^2 = 1$. (a)  $\alpha = 10^{-3}$ and $\omega/(2M_\Omega) \simeq 0.94818$. (b) $\alpha = 0.8$ and $\omega/(2M_\Omega) \simeq 0.93418$. (Online version in colour.)}
    \label{fig:IM2}
\end{figure}

We now outline the key properties of the non-toroidal inertial modes. 
We show in figure \ref{fig:IM1}b the effects of compressibility on the inertial spectrum, computed at $n=20$. 
Compressibility does not significantly modify the incompressible frequencies $\omega_0$ when $\alpha \ll 10^{-1}$, with variations $\Delta \omega = |\omega - \omega_0|/|\omega_0|$ smaller than one percent in figure \ref{fig:IM1}b. 
When $\alpha \to 1$, most of the compressible values depart by a few percent from the incompressible predictions, as observed for the highest-frequency modes. 
The meridional structure of the high-frequency inertial modes is also barely modified by compressibility, as illustrated in figure \ref{fig:IM2} for the kinetic energy.
The latter is maximum near the surface, which agrees with incompressible \cite{vantieghem2014inertial,zhang2017theory} and anelastic \cite{busse2005slow,wu2005origin} diffusionless predictions, and the structures remain similar to those of an incompressible homogeneous model with $\alpha=0$ (as previously reported for Earth-like conditions \cite{seyed2007inertial}).

\begin{figure}
    \centering
    \begin{tabular}{cc}
    \includegraphics[trim={1cm 4cm 1cm 4cm},clip,width=0.48\textwidth]{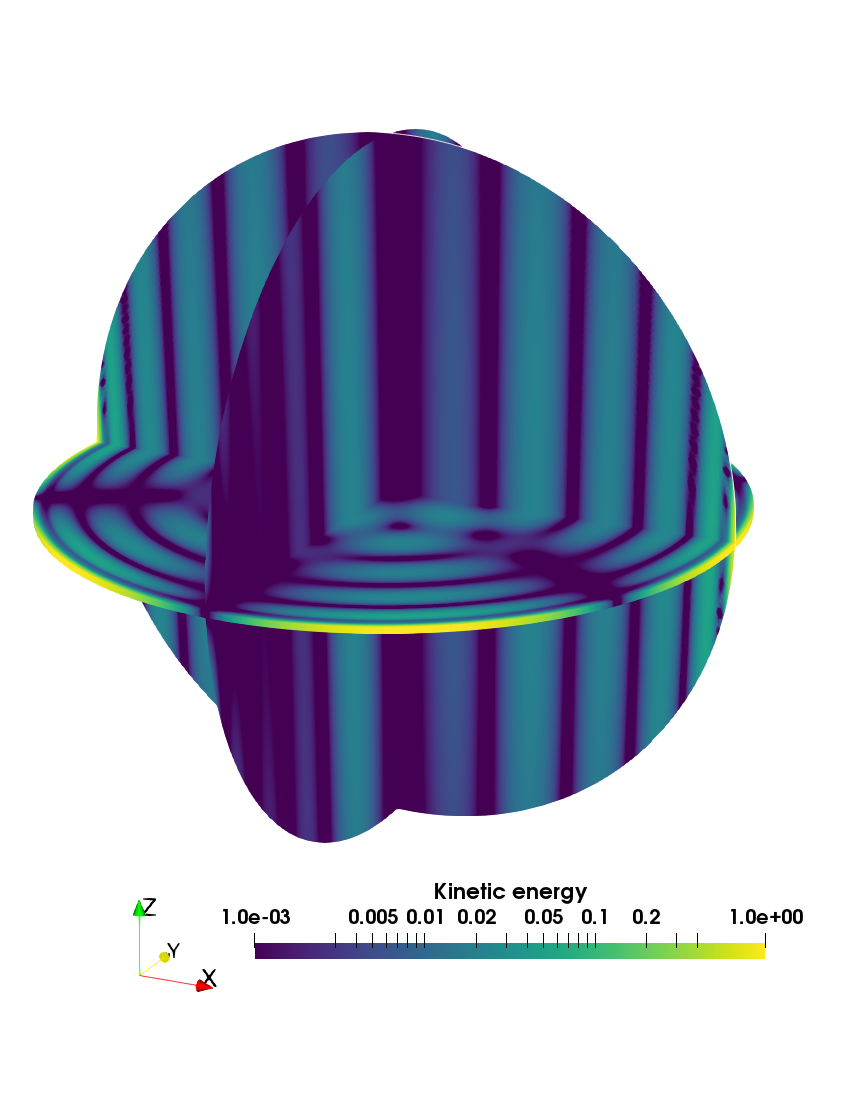} &
    \includegraphics[trim={1cm 4cm 1cm 4cm},clip,width=0.48\textwidth]{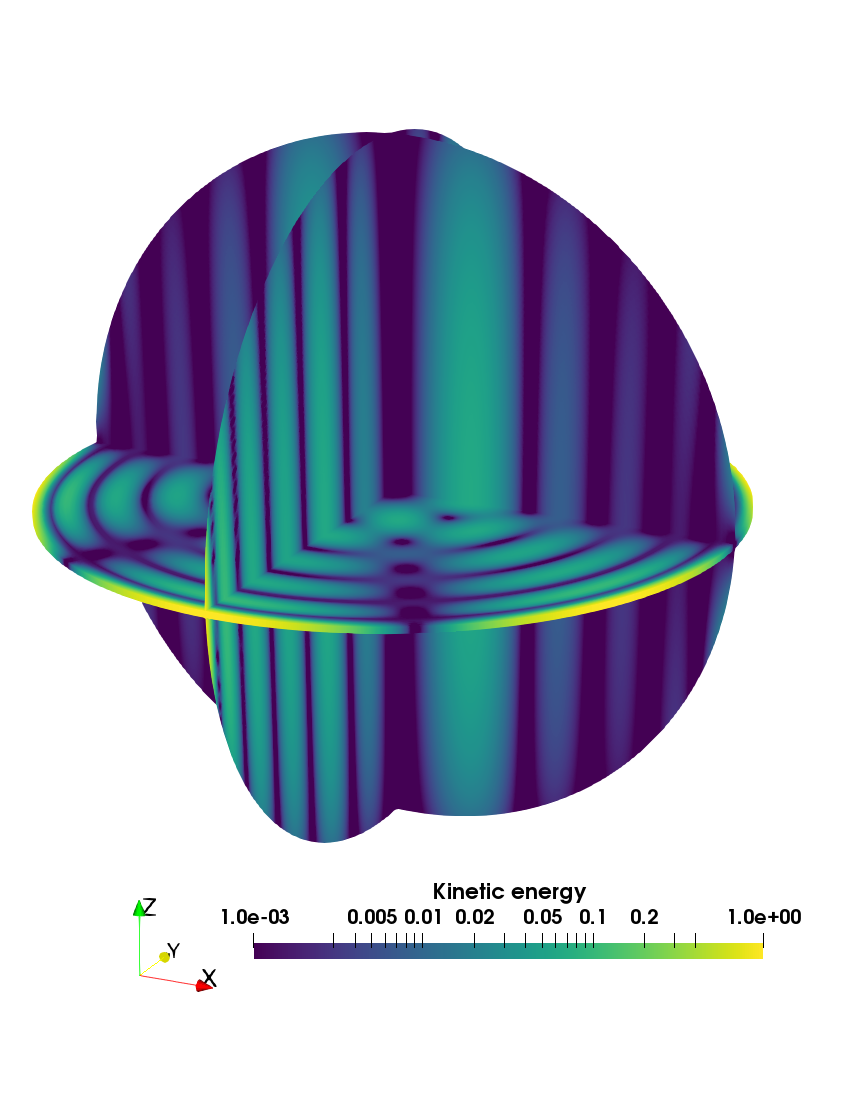} \\
    (a) & (b) \\
    \end{tabular}
    \caption{Quasi-geostrophic inertial mode in the sphere with $M_\Omega=10^{-2}$. Three-dimensional rendering of the kinetic energy $\rho_0 ||\boldsymbol{\zeta}||^2$. Colour bar is normalised such that $\max \rho_0 ||\boldsymbol{\zeta}||^2 = 1$. (a) $\alpha = 0.205$ and $\omega/(2M_\Omega) \simeq 0.01954$. (b) $\alpha = 0.8$ and $\omega/(2M_\Omega) \simeq 0.03266$. The modes have a different phase in (a) and (b). (Online version in colour.)}
    \label{fig:IM3}
\end{figure}

\begin{figure}
    \centering
    \begin{tabular}{cc}
    \includegraphics[trim={4cm 4cm 2cm 8.5cm},clip,width=0.48\textwidth]{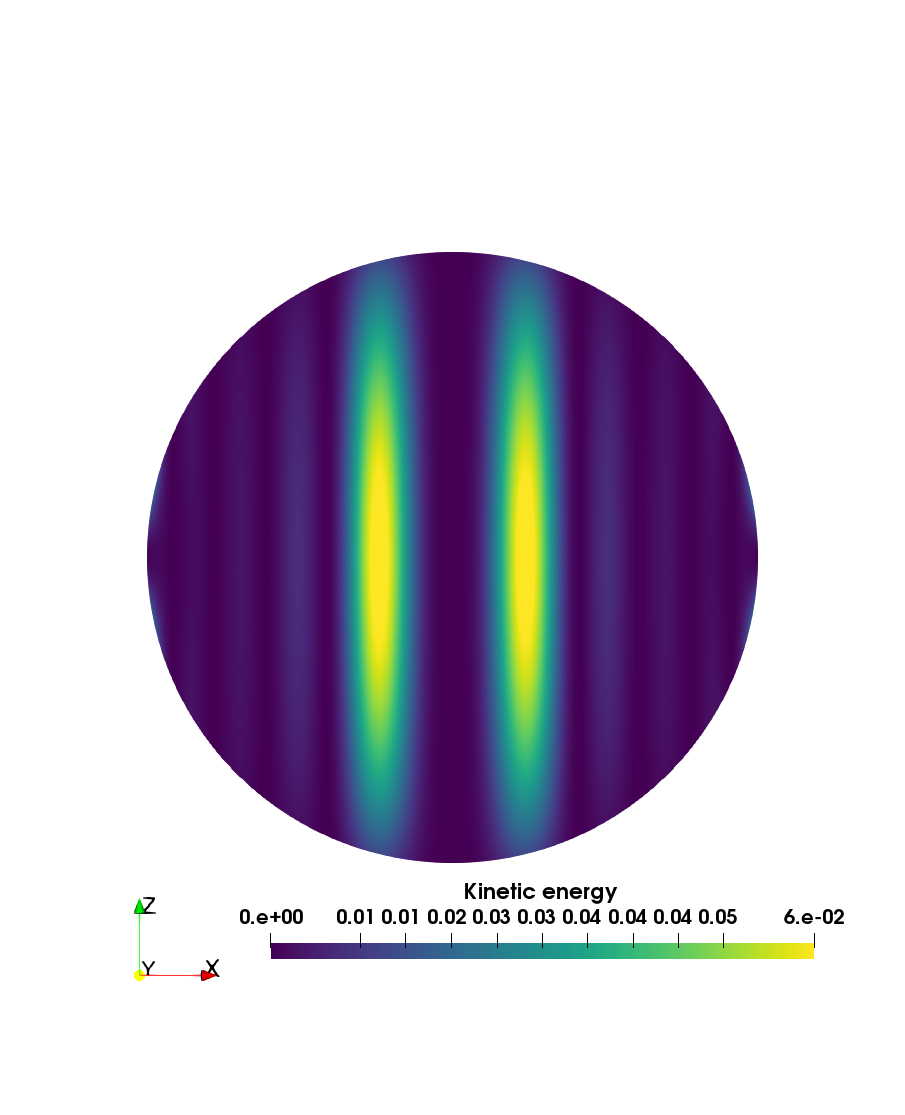} &
    \includegraphics[width=0.48\textwidth]{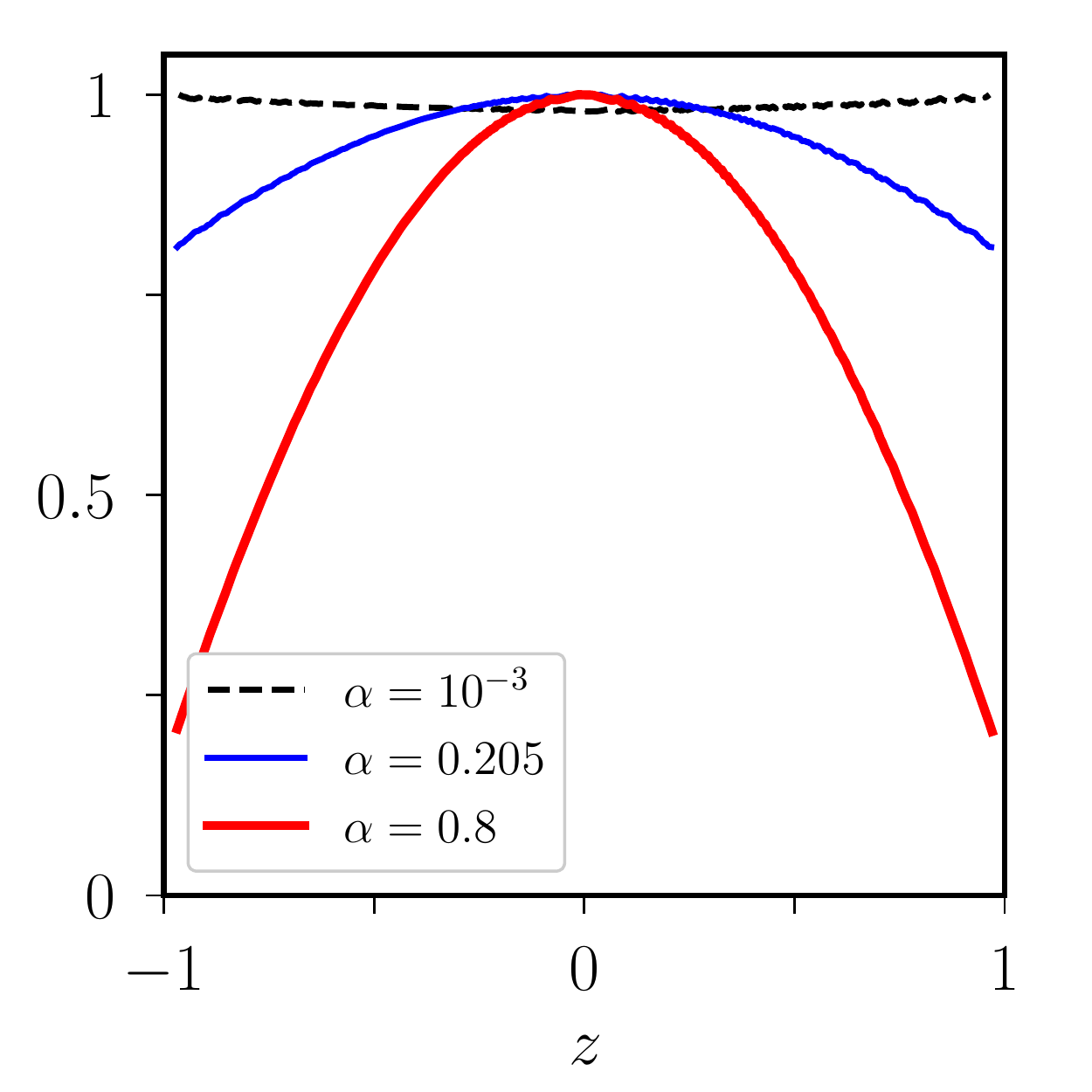} \\
    (a) & (b) \\
    \end{tabular}
    \caption{Evolution of quasi-geostrophy as a function of compressibility for the mode in figure \ref{fig:IM3}. (a) Meridional slice of figure \ref{fig:IM3}b, but with a linear scale. The one-dimensional (1-D) profiles have been measured along the most energetic QG column, at cylindrical radius $s=0.25$. (b) Evolution of the normalized kinetic energy $\rho_0 ||\boldsymbol{\zeta}||^2$ (normalised by its maximum value along the 1-D profile), as a function of $z$ for different values of $\alpha$. (Online version in colour.)}
    \label{fig:IM4}
\end{figure}

However, the lowest-frequency inertial modes are more affected by compressibility.
Discrepancies as large as $10\%$ are indeed obtained for $\Delta \omega$ with $\alpha=0.205$ and $M_\Omega=10^{-2}$ in figure \ref{fig:IM1}b, that is for Earth-like conditions.  
The lowest-frequency inertial modes consist of quasi-geostrophic (QG) modes \cite{maffei2017characterization}, which tend to the geostrophic limit when $\omega \to 0$. 
The geostrophic (zero-frequency) modes satisfy in isentropic interiors \cite{valette1989etude}
\begin{subequations}
\label{eq:kerT}
\begin{equation}
    \boldsymbol{\nabla} \boldsymbol{\cdot} (\rho_0 \, \boldsymbol{\zeta}) = 0, \quad \boldsymbol{\nabla} \times (\boldsymbol{1}_z \times \boldsymbol{\zeta}) = \boldsymbol{0}, \quad \boldsymbol{\zeta} \boldsymbol{\cdot} \boldsymbol{1}_n = 0 \ \, \text{on} \ \, \partial V.
    \tag{\theequation a--c}
\end{equation}
\end{subequations}
The geostrophic modes obey the modified Proudman-Taylor theorem $\partial \boldsymbol{\zeta}/\partial z = (\boldsymbol{\nabla} \boldsymbol{\cdot} \boldsymbol{\zeta}) \, \boldsymbol{1}_z$ in isentropic interiors, such that the equatorial geostrophic components are invariant along the axis of rotation (because $\boldsymbol{\nabla} \boldsymbol{\cdot} \boldsymbol{\zeta} \neq 0$). 
Thus, only the horizontal components $[\zeta_x, \zeta_y]$ of the QG modes could be almost invariant along the axis of rotation. 
The evolution of the spatial structure of an illustrative QG mode is shown in figure \ref{fig:IM3}, for mildly and strongly compressible models.  
The energy of this particular mode is maximum in the equatorial region, and actually dominates the energy of the other structures.
Thus, we have used a logarithmic scale to be able to observe all the patterns in the bulk.  
The domain is filled with columnar structures, which exhibit variations along the axis of rotation when compressibility is increased. 
This is more quantitatively evidenced by using a linear scale in figure \ref{fig:IM4}.
The latter shows that the energy $\rho_0 ||\boldsymbol{\zeta}||^2$ along the QG columns, which is dominated by $\rho_0 (\zeta_x^2 +\zeta_y^2)$, is reduced when $\alpha$ is increased (which is in part due to the geometrical factor $\rho_0$, not shown). 
The QG modes are thus modified by compressibility both in frequency and in energy (but the horizontal components of $\boldsymbol{\zeta}$ could be weakly modified in the volume).

\section{Planetary implications}
\label{sec:discuss}
The polynomial method has been used to compute the normal modes, considering here an idealized model of planetary interiors (as a proof-of-concept).  
Other reference states could be also considered (see the electronic supplementary material), but our results may have already implications for planetary models. 
We have notably shown that compressibility does alter the lowest-frequency QG modes, which are for instance often invoked to explain some features of the Earth's core dynamics \cite{bardsley2018could,gerick2020mhd}. 
Thus, it may be preferable to use the compressible modes to get more realistic physical insights into the planetary flow dynamics. 
To do so, we could consider the dynamical problem
\begin{equation}
   \frac{\partial^2 \boldsymbol{\xi}_1}{\partial t^2} +  \boldsymbol{\mathcal{C}} \left ( \frac{\partial \boldsymbol{\xi}_1}{\partial t} \right ) + \boldsymbol{\mathcal{K}} \left ( \boldsymbol{\xi}_1 \right ) = \boldsymbol{f} + \boldsymbol{\mathcal{N}}, \quad \boldsymbol{\xi}_1 \boldsymbol{\cdot} \boldsymbol{1}_n = 0 \ \, \text{on} \ \, \partial V,
   \label{eq:dynamicalpb}
\end{equation}
where $[\boldsymbol{\mathcal{C}}, \boldsymbol{\mathcal{K}}]$ are the linear operators given in \S\ref{sec:modalpb}\ref{subsec:infinite}, 
$\boldsymbol{f}$ represents external forces (e.g. tides \cite{ivanov2013unified}), and $\boldsymbol{\mathcal{N}}$ is a coupling operator including linearised advection terms \cite{lynden1967stability} (acting on $\boldsymbol{\xi}_1$ and its time derivative), as in the presence of large-scale tidally driven flows \cite{le2015flows}, and possibly nonlinear interactions with a second-order theory for the Lagrangian displacement \cite{roberts1967introduction,schutz1978langrangian}. 
Then, dynamical problem (\ref{eq:dynamicalpb}) could be converted into a finite-dimensional one in seeking
\begin{equation}
	\boldsymbol{\xi}_1 (\boldsymbol{r},t) = \sum_{j} \left [ \gamma_{j} (t) \, \boldsymbol{\zeta}_j \, (\boldsymbol{r}) + \gamma_{j}^{\dagger} (t) \, \boldsymbol{\zeta}_j^{\dagger} \, (\boldsymbol{r}) \right ], \quad \boldsymbol{\zeta}_j \boldsymbol{\cdot} \boldsymbol{1}_n = 0 \ \, \text{on} \ \, \partial V,
	\label{eq:expandweaklyNL}
\end{equation}
where the sum goes over a subset of complex-valued eigenmodes $\{ \boldsymbol{\zeta}_j \}$ of QEP (\ref{eq:compressibleQEP}). 
Modal expansions have been used in triaxial ellipsoids \cite{vantieghem2015latitudinal}, because expansion (\ref{eq:expandweaklyNL}) is exact when an infinite number of incompressible inertial modes is considered \cite{backus2017completeness,ivers2017enumeration}. 
Finally, we could circumvent the (daunting) computations to solve problem (\ref{eq:dynamicalpb}) by using the modal symmetries.

Two rotating compressible eigen-pairs $[\omega_1, \boldsymbol{\zeta}_1]$ and $[-\omega_2, \boldsymbol{\zeta}_2^{\dagger}]$ with $\omega_1\neq \omega_2$ satisfy
\begin{subequations}
\label{eq:prooforthogonality}
\begin{equation}
    \omega_1^2 \rho_0 \, \boldsymbol{\zeta}_1 = \mathrm{i} \omega_1 \, \rho_0 \, \boldsymbol{\mathcal{C}} (\boldsymbol{\zeta}_1) + \rho_0 \, \boldsymbol{\mathcal{K}} (\boldsymbol{\zeta}_1), \quad
    \omega_2^2 \, \rho_0 \, \boldsymbol{\zeta}_2^{\dagger} = - \mathrm{i} \omega_2 \, \rho_0 \, \boldsymbol{\mathcal{C}} (\boldsymbol{\zeta}_2^{\dagger}) + \rho_0 \, \boldsymbol{\mathcal{K}}(\boldsymbol{\zeta}_2^{\dagger}).
    \tag{\theequation a,b}
\end{equation}
\end{subequations}
Taking the dot product of the first equality with $\boldsymbol{\zeta}_2^{\dagger}$ and of the second one with $\boldsymbol{\zeta}_1$, we obtain the modified orthogonality condition (by virtue of the symmetries of $\boldsymbol{\mathcal{C}}$ and $\boldsymbol{\mathcal{K}}$)
\begin{equation}
    \left ( \omega_1 + \omega_2 \right ) \, \langle \boldsymbol{\zeta}_2, \boldsymbol{\zeta}_1 \rangle_{\rho_0} = \langle \boldsymbol{\zeta}_2, \mathrm{i} \boldsymbol{\mathcal{C}}(\boldsymbol{\zeta}_1) \rangle_{\rho_0} .
    \label{eq:orthogonality1}
\end{equation}
Then, we combine (\ref{eq:orthogonality1}) and (\ref{eq:prooforthogonality}) to obtain another modified orthogonality condition
\begin{equation}
    \omega_1 \, \omega_2 \, \langle \boldsymbol{\zeta}_2, \boldsymbol{\zeta}_1 \rangle_{\rho_0} = - \langle \boldsymbol{\zeta}_2, \boldsymbol{\mathcal{K}}(\boldsymbol{\zeta}_1) \rangle_{\rho_0}.
    \label{eq:orthogonality2}
\end{equation}
The rotating compressible modes are thus not orthogonal with respect to inner product (\ref{eq:scalprod0}), but satisfy the modified orthogonality relations (\ref{eq:orthogonality1}) and (\ref{eq:orthogonality2}). 
The latter conditions are even more general, because they only result from the symmetries of the QEP \cite[for finite-dimensional problems]{barston1967eigenvalue}. 
Note that the orthogonality condition of the non-rotating compressible modes is also recovered from condition (\ref{eq:orthogonality1}) when $\boldsymbol{\mathcal{C}}=\boldsymbol{0}$. 
Therefore, it may be preferable to use the above orthogonality conditions to project the dynamical equations onto a subset of normal modes. 

Actually, the reduction of (\ref{eq:dynamicalpb}) onto a finite-dimensional problem using the modified orthogonality conditions is underpinned by the mathematical properties of the operator
\begin{equation}
    \boldsymbol{\mathcal{T}} = \begin{pmatrix}
		0 & 1 \\
		-\boldsymbol{\mathcal{K}} & - \boldsymbol{\mathcal{C}}\\
	\end{pmatrix},
	\label{eq:GEP_infinite}
\end{equation}
which appears in the left-hand side of problem (\ref{eq:dynamicalpb}). 
A rigorous analysis \cite[proposition 5]{valette1989etude} reveals that the zero-frequency modes, so the geostrophic modes in rotating isentropic interiors, are associated with Jordan chains in QEP (\ref{eq:QEPfinite}). 
Hence, only modal expansions (\ref{eq:expandweaklyNL}) that exclude any geostrophic solutions could be used in association with the modified orthogonality conditions \cite{wahr1981normal}. 
Including the Jordan-chain modes, 
which obey different modified orthogonality conditions \cite{schutz1980perturbations}, would severely complicate the mathematical formulation of the reduced problem. 
Thus, the use of modal expansion (\ref{eq:expandweaklyNL}) to solve problem (\ref{eq:dynamicalpb}) does not appear to be of great computational interest in the presence of geostrophic modes.   

However, the generation of geostrophic motions is a long-standing issue in the theory of rotating fluids even in the incompressible regime  \cite{greenspan1969non}. 
The observed zonal winds at Jupiter's surface could be also related to deep geostrophic flows \cite{busse1976simple,glatzmaier2018computer}, but different physical scenarios are highly disputed in light of recent high-precision data \cite{guillot2018suppression,moore2019time,christensen2020mechanisms}. 
To assess the robustness of the deep scenario, the polynomial formulation might be used directly to solve dynamical problem (\ref{eq:dynamicalpb}) without any prior computations of the normal modes (as for incompressible theories \cite{lebovitz1996new,vidal2017inviscid}).
This might allow us to circumvent the mathematical difficulty associated with the projection onto the geostrophic modes. 
Clarifying the usefulness of the polynomial method for that problem deserves future work. 
Moreover, to investigate the longer-term nonlinear dynamics, we should also reintroduce viscosity to smooth out spurious small scales that could develop over time. 
A preliminary asymptotic viscous theory is presented in appendix \ref{appendix:viscous}.

\section{Concluding remarks}
\label{sec:ccl}
We have investigated the normal modes in rotating fluid-filled ellipsoids, as an idealized model of isentropic planetary interiors. 
Since the various sound-proof approximations are still debated, we have rigorously attacked the fully compressible diffusionless problem and considered the coreless ellipsoidal geometry, to avoid any mathematical singularities for zero viscosity (contrary to shells). 
We have developed a spectral algorithm to sidestep the mathematical complexity of the ellipsoidal coordinates, combining a global polynomial description made of polynomial elements in the Cartesian coordinates \cite{vidal2020acoustics}, applied here to the weighted Helmholtz decomposition, and a Galerkin (projection) method pioneered by Lebovitz \cite{lebovitz1989stability}. 
We have used the method to numerically investigate the linear eigenvalue problem, formulated for the Lagrangian displacement. 
We have thoroughly assessed the accuracy of the polynomial solutions, against targeted finite-element computations (or analytical predictions).

We have first considered the acoustic modes, which are mildly modified by compressibility.
We have notably explored the effects of the ellipticity and global rotation, showing that first-order perturbative computations are not accurate enough to predict the modal angular frequencies in rapidly rotating planets (e.g. gas giants, but also for stars \cite{reese2006acoustic}). 
We have then investigated the inertial modes, showing that mainly the low-frequency (quasi-geostrophic) modes are affected by compressibility. 
We have finally discussed the usefulness of the normal modes to build reduced models of planetary flow dynamics. 
Modal expansions are certainly appropriate to investigate flow instabilities generated in the presence of large-scale orbitally driven flows, but their practical interest appears unsuitable for describing the generation of geostrophic flows (with dynamical models for the displacement vector).  

Several physical issues have remained unaddressed in this work. 
We have removed the viscosity from the modal problem, but we have presented a preliminary asymptotic theory to reintroduce a posteriori the viscous effects for the stress-free conditions. 
Yet, the latter conditions may appear too severe for planetary liquid cores, which are surrounded by solid mantles.
Indeed, stress-free conditions filter out the Ekman boundary layer that should exist at the core-mantle boundary. 
Additional viscous effects that affect the inviscid modes in the coreless ellipsoid have also been neglected, such as the eruption of boundary layers at critical latitudes \cite{stewartson1963motion}. 
They are however expected to be more prominent for forced dynamical problems \cite{lin2015shear}, and may even only barely modify the global viscous damping of the modes (as found in the incompressible case \cite{hollerbach1995oscillatory}).
Therefore, it may be possible to parametrize the surface viscous damping of the modes (as for the incompressible modes \cite{liao2001viscous,liao2010asymptotic,lemasquerier2017libration}). 
Using ellipsoidal coordinates would be necessary to solve the compressible boundary layer equations (by analogy with compressible spherical computations \cite{abney1996ekman,glampedakis2006ekman}), which is a challenging problem. 
For gas giants, it might be possible to include self-gravitation for polynomial density profiles \cite{lebovitz1979ellipsoidal}, and consider non-rigid boundaries \cite{lebovitz1989stability,vidal2019acoustics} (but only with a non-zero density on the boundary).  

Direct applications of our work include investigating the elliptical instability \cite{kerswell2002elliptical} in fully compressible ellipsoids \cite{cebron2013elliptical}. 
We could start with the hydrostatic state we have presented here, and then other hydrostatic states could be considered (for an improved planetary accuracy).  
We could also investigate the normal modes in stably stratified rotating interiors. 
The polynomial method is probably not well adapted to compute the highest-frequency (acoustic or gravity) modes, which exhibit localised structures that are already well described by ray theory \cite{lignieres2009asymptotic,prat2016asymptotic}.
Yet, it could be used to compute the lowest-frequency inertial-gravity modes \cite{friedlander1982internal} that can have larger-scale components \cite[in spheres]{seyed2015effects}. 
Moreover, the inertial-gravity modes are known to be the preferred modes for nonlinear couplings with tidal flows in rotating stratified interiors \cite{kerswell1993elliptical,vidal2019binaries}, but the Boussinesq results remain to be extended with compressibility. 
We hope the present study will shed light on the polynomial method in fluid ellipsoids.

\enlargethispage{20pt}


\dataccess{The paper has \href{https://doi.org/10.6084/m9.figshare.c.5056742}{electronic supplementary material}, including the supporting data for most of the figures. The source code \textsc{shine} is released at \url{https://bitbucket.org/vidalje/shine/}.}

\aucontribute{This work is an original idea of J.V., who designed the study, developed the mathematical analysis and performed the numerical computations. D.C. conducted the finite-element computations to benchmark the polynomial method. J.V. and D.C. discussed and approved the results presented in the article. J.V. drafted the paper and both authors gave final approval for submission.}

\competing{The authors declare that they have no competing interests.}

\funding{D.C. has received funding from the European Research Council (ERC) under the European Union's Horizon 2020 research and innovation programme (grant agreement No 847433).}

\ack{The authors acknowledge the two anonymous reviewers for their comments that improved the quality of the original manuscript.
J.V. is also grateful to Dr B. Valette for valuable discussions. 
The spherical harmonic decompositions have been done using the open-source library \textsc{shtns} \cite{schaeffer2013efficient}.
Figures were produced using matplotlib (\url{http://matplotlib.org/}) and paraview (\url{http://www.paraview.org/}). 
}

\appendix
\numberwithin{equation}{section}

\section{Spectral decompositions in rigid ellipsoids}
\label{appendix:spectral}
We elaborate on the spectral decompositions in rigid triaxial ellipsoids. 
We give the gauge conditions to properly define the weighed Helmholtz decomposition in \S \ref{appendix:spectral}\ref{appendix:spectral1}, then we provide the admissible polynomial forms in \S \ref{appendix:spectral}\ref{appendix:basis}, and finally we outline the explicit solutions of the gauge conditions in \S \ref{appendix:spectral}\ref{appendix:spectral2}. 

\subsection{Gauge conditions}
\label{appendix:spectral1}
We start with Helmholtz decomposition (\ref{eq:helmholtzweighted2}), which expresses $\rho_0 \boldsymbol{\zeta}$ in terms of two sub-spaces that are not mutually orthogonal with respect to inner product (\ref{eq:weighedinnerprodcuct}). 
To partition the vector space into orthogonal sub-spaces, we use the gauge transformation \cite{sobouti1981potentials} 
\begin{subequations}
\label{eq:gaugetransformation}
\begin{equation}
  \widehat{\Phi} = \rho_0 \Phi + \Phi_1, \quad \widehat{\boldsymbol{\Psi}} = \boldsymbol{\Psi} + \boldsymbol{\Psi}_1. 
    \tag{\theequation a,b}
\end{equation}
\end{subequations}
The weighted Helmholtz decomposition (\ref{eq:helmholtzweighted}) is recovered from decomposition (\ref{eq:helmholtzweighted2}) under the gauge equations for the vector potential 
\begin{subequations}
\label{eq:gauge1}
\begin{equation}
    \boldsymbol{\nabla} \times \boldsymbol{\Psi}_1 = -\Phi \, \boldsymbol{\nabla} \rho_0  - \boldsymbol{\nabla} \Phi_1, \quad  (\boldsymbol{\nabla} \times \boldsymbol{\Psi}_1) \boldsymbol{\cdot} \boldsymbol{1}_n = 0 \ \, \text{on} \ \, \partial V,
    \tag{\theequation a,b}
\end{equation}
\end{subequations}
and the scalar potential
\begin{subequations}
\label{eq:gauge2}
\begin{equation}
    \nabla^2 \Phi_1 = - \boldsymbol{\nabla} \boldsymbol{\cdot} (\Phi \, \boldsymbol{\nabla} \rho_0), \quad (\boldsymbol{\nabla} \Phi_1) \boldsymbol{\cdot} \boldsymbol{1}_n = - (\Phi \, \boldsymbol{\nabla} \rho_0) \boldsymbol{\cdot} \boldsymbol{1}_n \ \, \text{on} \ \, \partial V.
    \tag{\theequation a,b}
\end{equation}
\end{subequations}
Equations (\ref{eq:gauge1})-(\ref{eq:gauge2}) admit solutions for $[\boldsymbol{\Psi}_1, \Phi_1]$ in rigid ellipsoids (see below), such that decompositions (\ref{eq:helmholtzweighted}) and (\ref{eq:helmholtzweighted2}) are equivalent. 

Two limit situations are worth discussing. 
The acoustic modes are such that $\boldsymbol{\xi} = \nabla \Phi$ in non-rotating isentropic interiors \cite{sobouti1981potentials}. 
This implies $\Phi_1 \neq 0$ from the gauge equation (\ref{eq:gauge2}), which then leads to $\boldsymbol{\Psi}_1 \neq \boldsymbol{0}$ from (\ref{eq:gauge1}), and so $\widehat{\boldsymbol{\Psi}} \neq \boldsymbol{0}$.
Therefore, the acoustic modes are described by the two non-vanishing potentials $[\widehat{\Phi}, \widehat{\boldsymbol{\Psi}}]$ 
in decomposition (\ref{eq:helmholtzweighted2}). 
Conversely, purely anelastic flows (with $\Phi=0$ and $\boldsymbol{\Psi} \neq \boldsymbol{0}$) are only described by the vector potential $\widehat{\boldsymbol{\Psi}} \neq \boldsymbol{0}$ in decomposition (\ref{eq:helmholtzweighted2}). 

\subsection{Basis elements}
\label{appendix:basis}
We present admissible polynomial basis elements for the finite-dimensional spaces  $\boldsymbol{\mathcal{V}} \, [n\geq1]$ and $\boldsymbol{\mathcal{W}} \, [n\geq2]$ \cite{vidal2020acoustics}.
We introduce the finite-dimensional space $\mathcal{P} \,[n]$, which is spanned by the scalar monomials $x^iy^jz^k$ with $i+j+k \leq n$. 
To construct the elements in $\boldsymbol{\mathcal{V}} \, [n\geq1]$, we consider the linearly independent Cartesian monomials in $\mathcal{P} \,[n-1]$.
Their number is $N_2 = n (n+1) (n+2)/6$. 
Among them, there are $N_1 = n (n+1)/2$ monomials that are independent of $z$, denoted $\{\mathfrak{g}_i\}$. 
The other monomials, denoted $\{\mathfrak{h}_j\}$, contain $z$ and its powers as factor. 
These polynomials are indexed with $(i,j) \in [1, N_1] \times [N_1 + 1, N_2]$ as
\begin{subequations}
\label{Eq_ghi}
\begin{equation}
	\left \{ \mathfrak{g}_i \right \} = \left \{ 1,x,y,x^2,xy,y^2,\dots, x^{n-1}, y^{n-1} \right \}, \quad
	\left \{ \mathfrak{h}_j \right \} = \left \{ z, xz, yz, z^2, \dots, z^{n-1} \right \}, 
    \tag{\theequation a,b}
\end{equation}
\end{subequations}
and we denote $\{ \mathfrak{p}_k\}_{k \leq N_2} = \{\mathfrak{g}_i\} \, \bigcup \, \{\mathfrak{h}_j\}$. 
Admissible elements $\boldsymbol{e} \in \boldsymbol{\mathcal{V}} \, [n\geq1]$ are then \cite{lebovitz1989stability}
\begin{subequations}
\label{eq:BasisVn}
\begin{equation}
\boldsymbol{e}_{k} = \boldsymbol{\nabla} [\mathfrak{p}_k \, (F-1) ] \times \boldsymbol{1}_x, \ \boldsymbol{e}_{N_2 + k} = \boldsymbol{\nabla} [\mathfrak{p}_k \, (F-1) ] \times \boldsymbol{1}_y, \ \boldsymbol{e}_{2N_2+i} = \boldsymbol{\nabla} [\mathfrak{g}_i \, (F-1) ] \times \boldsymbol{1}_z, 
\tag{\theequation a--c}
\end{equation}
\end{subequations}
with $1 \leq k \leq N_2$, $1 \leq i \leq N_1$, and the shape function $F = (x/a)^{2} + (y/b)^{2} + (z/c)^{2}$. 
We have $\dim \boldsymbol{\mathcal{V}} \, [n\geq1] = N_1 + 2N_2 = n (n+1)(2n+7)/6$. 
To construct basis elements for $\boldsymbol{\mathcal{W}} \, [n\geq2]$, we introduce the operator $\mathcal{N}^{-1}$ defined as \cite{vidal2020acoustics}
\begin{equation}
	\mathcal{N}^{-1} \{x^i y^j z^k \} = \frac{1}{(i/a^2 + j/b^2 + k/c^2)^2} \, \left ( \frac{x}{a^2} \frac{\partial}{\partial x} + \frac{y}{b^2} \frac{\partial}{\partial y} + \frac{z}{c^2} \frac{\partial}{\partial z} \right ) x^i y^j z^k, 
	\label{eq:Ninv}
\end{equation}
for any monomial $x^i y^j z^k$ with $i+j+k \geq 1$. 
Then, basis elements $\boldsymbol{e} \in \boldsymbol{\mathcal{W}} \, [n\geq2]$ are given by
\begin{subequations}
\label{eq:elementsWn}
\begin{equation}
	\boldsymbol{e} = \boldsymbol{\nabla} \widehat{\Phi}, \quad \widehat{\Phi} = \mathcal{N}^{-1} \left \{ ( F-1) \, \Psi \right \},
    \tag{\theequation a,b}
\end{equation}
\end{subequations}
with the scalar polynomial $\Psi \in \mathcal{P} \, [1,n-1]$. 
We get $\dim \boldsymbol{\mathcal{W}} \, [n \geq 2] = n(n+1)(n+2)/6 - 1$.

\subsection{Admissible solutions of the gauge equations}
\label{appendix:spectral2}
Solutions of the gauge equations (\ref{eq:gauge1})-(\ref{eq:gauge2}) can be found as follows. 
We consider a polynomial density profile $\rho_0$, made of Cartesian monomials of maximum degree $n_{\rho}$. 
The vector field $\rho_0 \boldsymbol{\nabla} \Phi$ in decomposition (\ref{eq:helmholtzweighted}) possesses admissible polynomial forms of maximum degree $n$ belonging to $\boldsymbol{\mathcal{W}} \, [n\geq2]$, such that $\Phi$ is a polynomial of maximum degree $n+1-n_{\rho}$. 
Similarly, $\boldsymbol{\nabla} \times \boldsymbol{\Psi}$ can be seek in $\boldsymbol{\mathcal{V}} \, [n\geq1]$ where $\boldsymbol{\Psi}$ is a polynomial vector of maximum degree $n+1$. 
To satisfy the inhomogeneous Neumann condition (\ref{eq:gauge2}b), the solution of gauge equation (\ref{eq:gauge2}) can be written as the sum $\Phi_1 = \Phi_1^H + \Phi_1^P$ made of two polynomial solutions of maximum degree $n+1$.  
The particular solution $\Phi_1^P$, which satisfies the boundary condition (\ref{eq:gauge2}b), is given by \cite[p. 358]{cartan1922petites}
\begin{equation}
    \Phi_1^P = - \mathcal{N}^{-1} \left ( \Phi \, \boldsymbol{\nabla} \rho_0  \boldsymbol{\cdot} \boldsymbol{n} \right ) \quad \text{with} \quad \boldsymbol{n} = (x/a^2, y/b^2, z/c^2)^\top.
    \label{eq:up}
\end{equation}
Then, we have $\nabla^2 \Phi_1^H = - \nabla^2 \Phi_1^P - \boldsymbol{\nabla} \boldsymbol{\cdot} (\Phi \, \boldsymbol{\nabla} \rho_0)$ together with the boundary condition $\boldsymbol{\nabla} \Phi_1^H \boldsymbol{\cdot} \boldsymbol{1}_n = 0$, such that we can seek $\boldsymbol{\nabla} \Phi_1^H \in \boldsymbol{\mathcal{W}} \, [n\geq2]$. 
Then, we can obtain a polynomial expression of maximum degree $n+1$ for $\widehat{\Phi}$ from transformation (\ref{eq:gaugetransformation}a). 
Finally, $\boldsymbol{\nabla} \times \boldsymbol{\Psi}_1$ defined in (\ref{eq:gauge1}a) admits a polynomial expression of maximum degree $n$ and belonging to $\boldsymbol{\mathcal{V}} \, [n\geq1]$, since it satisfies the non-penetration boundary condition. 

\section{Preliminary viscous theory for stress-free conditions}
\label{appendix:viscous}
Viscous boundary conditions cannot be rigorously enforced with the polynomial method (because of the global nature of the basis elements), and so perturbation theory must be employed to model viscous effects in the planetary limit $Ek\to 0$. 
We present an asymptotic theory for the stress-free conditions, which is compatible with the polynomial formulation.  
Indeed, perturbation theory for the stress-free conditions does not require the spatial structure of the viscous flows (contrary to the no-slip conditions), as first uncovered in incompressible spheres \cite{liao2001viscous}.  

We work in dimensional units but, exceptionally, we omit here the superscripts ${}^{\ast}$ to simplify the expressions. 
The linearised viscous force, which ought to be included in the right-hand side of equation (\ref{eq:dynamicalpb}) for the displacement, is given by
\begin{subequations}
\label{eq:viscousforcexi}
\begin{equation}
    \boldsymbol{\mathcal{F}}_\eta = (1/\rho_0) \, \boldsymbol{\nabla} \boldsymbol{\cdot} [\eta \, \boldsymbol{\sigma} (\dot{\boldsymbol{\xi}}_1 + \dot{\bar{\boldsymbol{\xi}}}_1)], \quad \boldsymbol{\sigma} (\boldsymbol{a}) = \boldsymbol{\nabla} \boldsymbol{a} + (\boldsymbol{\nabla} \boldsymbol{a})^\top - ({2}/{3}) \, (\boldsymbol{\nabla} \boldsymbol{\cdot} \boldsymbol{a}) \, \boldsymbol{\mathcal{I}},
    \tag{\theequation a,b}
\end{equation}
\end{subequations}
where $\boldsymbol{\sigma}$ the strain rate tensor with a zero second viscosity (Stokes's hypothesis), $\boldsymbol{\mathcal{I}}$ is the identity tensor, $\eta = \rho_0 \, \nu$ is the dynamic (shear) viscosity, and $\bar{\boldsymbol{\xi}}_1$ is the leading-order viscous correction.
The latter allows the total displacement to satisfy the stress-free conditions (at leading order)
\begin{subequations}
\label{eq:SFBCxi}
\begin{equation}
\left [ \boldsymbol{\xi}_1 , \bar{\boldsymbol{\xi}}_1 \right ] \boldsymbol{\cdot} \boldsymbol{1}_n = 0, 
\quad 
\left [ \boldsymbol{1}_n \boldsymbol{\cdot} \boldsymbol{\sigma} (\boldsymbol{\xi}_1 + \bar{\boldsymbol{\xi}}_1) \right ] \times \boldsymbol{1}_n = \boldsymbol{0}.
\tag{\theequation a,b}
\end{equation}
\end{subequations}
Then, to be consistent with the Galerkin formulation of the inviscid formulation, we project viscous force (\ref{eq:viscousforcexi}) onto $\rho_0\boldsymbol{\xi}_1 = \boldsymbol{e}$ and integrate by parts. 
We have
\begin{equation}
    \langle \boldsymbol{e}, \boldsymbol{\mathcal{F}}_\eta \rangle_{\rho_0} = \int_V \boldsymbol{e}^{\dagger} \boldsymbol{\cdot} \left ( \boldsymbol{\nabla} \boldsymbol{\cdot} [ \eta \, \boldsymbol{\sigma} (\dot{\boldsymbol{\xi}}_1 + \dot{\bar{\boldsymbol{\xi}}}_1)] \right ) \, \mathrm{d}V,
\end{equation}
which simplifies into \cite{guermond2013remarks}
\begin{subequations}
\begin{align}
    \langle \boldsymbol{e}, \boldsymbol{\mathcal{F}}_\eta \rangle_{\rho_0} &= - \int_V \eta \, \boldsymbol{\nabla} \boldsymbol{e}^{\dagger} \, :  \, \boldsymbol{\sigma} (\dot{\boldsymbol{\xi}}_1 + \dot{\bar{\boldsymbol{\xi}}}_1) \, \mathrm{d} V - \int_S \eta \left [ \boldsymbol{1}_n \boldsymbol{\cdot} \boldsymbol{\sigma} (\dot{\boldsymbol{\xi}}_1 + \dot{\bar{\boldsymbol{\xi}}}_1) \times \boldsymbol{1}_n  \right ] \boldsymbol{\cdot} (\boldsymbol{1}_n \times \boldsymbol{e}^{\dagger}) \, \mathrm{d} S, \label{eq:viscint1} \\
    &= - \int_V \eta \, \boldsymbol{\nabla} \boldsymbol{e}^{\dagger} \, : \, \boldsymbol{\sigma} (\dot{\boldsymbol{\xi}}_1 + \dot{\bar{\boldsymbol{\xi}}}_1) \, \mathrm{d} V \, = - (1/2) \int_V \eta \, \boldsymbol{\sigma} (\boldsymbol{e}^{\dagger}) \, : \, \boldsymbol{\sigma} (\dot{\boldsymbol{\xi}}_1 + \dot{\bar{\boldsymbol{\xi}}}_1) \, \mathrm{d} V, \label{eq:viscint2} \\
    &\simeq - (1/2) \int_V \eta \, \boldsymbol{\sigma} (\boldsymbol{e}^{\dagger}) \, : \, \boldsymbol{\sigma} (\dot{\boldsymbol{\xi}}_1) \, \mathrm{d} V, \label{eq:viscint3}
\end{align}
\end{subequations}
where $:$ denotes the contraction of the two tensors. 
The surface integral in equality (\ref{eq:viscint1}) vanishes by virtue of stress-free conditions (\ref{eq:SFBCxi}), and we have neglected $\boldsymbol{\sigma} (\dot{\bar{\boldsymbol{\xi}}}_1)$ in (\ref{eq:viscint3}) which is a second-order term. 
Viscous projection (\ref{eq:viscint3}), involving the total solution $\boldsymbol{\xi}_1$, possibly allows viscous couplings between the inviscid modes (at leading order). 
The latter may have important effects for the dynamics, but they remain to be quantitatively evaluated.



{
\bibliography{./main}

\begin{thebibliography}{99}

\bibitem{schaeffer2017turbulent}
Schaeffer N, Jault D, Nataf HC, Fournier A. 2017  Turbulent geodynamo
  simulations: a leap towards {Earth}'s core. {\em Geophys. J. Int.}
  \textbf{211}, 1--29.
(\href{http://dx.doi.org/doi:10.1093/gji/ggx265}{doi:10.1093/gji/ggx265}).

\bibitem{sheyko2018scale}
Sheyko A, Finlay C, Favre J, Jackson A. 2018  Scale separated low viscosity
  dynamos and dissipation within the {Earth's} core. {\em Sci. Rep.}
  \textbf{8}, 1--7.
(\href{https://doi.org/10.1038/s41598-018-30864-1}{doi:10.1038/s41598-018-30864-1}).

\bibitem{guervilly2019nature}
Guervilly C, Cardin P, Schaeffer N. 2019  Turbulent convective length scale in
  planetary cores. {\em Nature} \textbf{570}, 368--371.
(\href{https://dx.doi.org/10.1038/s41586-019-1301-5}{doi:10.1038/s41586-019-1301-5}).

\bibitem{chen2018optimal}
Chen L, Herreman W, Li K, Livermore PW, Luo JW, Jackson A. 2018  The optimal
  kinematic dynamo driven by steady flows in a sphere. {\em J. Fluid Mech.}
  \textbf{839}, 1--32.
(\href{https://dx.doi.org/10.1017/jfm.2017.924}{doi:10.1017/jfm.2017.924}).

\bibitem{daria2019dynamo}
Holdenried-Chernoff D, Chen L, Jackson A. 2019  A trio of simple optimized
  axisymmetric kinematic dynamos in a sphere. {\em Proc. R. Soc. A}
  \textbf{475}, 20190308.
(\href{https://dx.doi.org/10.1098/rspa.2019.0308}{doi:10.1098/rspa.2019.0308}).

\bibitem{kong2018origin}
Kong D, Zhang K, Schubert G, Anderson JD. 2018  Origin of {Jupiter}'s
  cloud-level zonal winds remains a puzzle even after {Juno}. {\em Proc. Natl
  Acad. Sci. USA} \textbf{115}, 8499--8504.
(\href{https://dx.doi.org/10.1073/pnas.1805927115}{doi:10.1073/pnas.1805927115}).

\bibitem{kloss2019time}
Kloss C, Finlay CC. 2019  Time-dependent low-latitude core flow and geomagnetic
  field acceleration pulses. {\em Geophys. J. Int.} \textbf{217}, 140--168.
(\href{https://dx.doi.org/10.1093/gji/ggy545}{doi:10.1093/gji/ggy545}).

\bibitem{greenspan1968theory}
Greenspan HP. 1968 {\em The theory of rotating fluids}.
Cambridge, UK: Cambridge University Press.

\bibitem{rieutord2000wave}
Rieutord M, Georgeot B, Valdettaro L. 2000  Wave attractors in rotating fluids:
  a paradigm for ill-posed {Cauchy} problems. {\em Phys. Rev. Lett.}
  \textbf{85}, 4277.
(\href{https://dx.doi.org/10.1103/PhysRevLett.85.4277}{doi:10.1103/PhysRevLett.85.4277}).

\bibitem{stewartson1969pathological}
Stewartson K, Rickard J. 1969  Pathological oscillations of a rotating fluid.
  {\em J. Fluid Mech.} \textbf{35}, 759--773.
(\href{https://dx.doi.org/10.1017/S002211206900142X}{doi:10.1017/S002211206900142X}).

\bibitem{rieutord2001inertial}
Rieutord M, Georgeot B, Valdettaro L. 2001  Inertial waves in a rotating
  spherical shell: attractors and asymptotic spectrum. {\em J. Fluid Mech.}
  \textbf{435}, 103--144.
(\href{https://dx.doi.org/10.1017/S0022112001003718}{doi:10.1017/S0022112001003718}).

\bibitem{rieutord2018axisymmetric}
Rieutord M, Valdettaro L. 2018  Axisymmetric inertial modes in a spherical
  shell at low {Ekman} numbers. {\em J. Fluid Mech.} \textbf{844}, 597--634.
(\href{https://dx.doi.org/10.1017/jfm.2018.201}{doi:10.1017/jfm.2018.201}).

\bibitem{backus2017completeness}
Backus G, Rieutord M. 2017  Completeness of inertial modes of an incompressible
  inviscid fluid in a corotating ellipsoid. {\em Phys. Rev. E} \textbf{95},
  053116.
(\href{https://dx.doi.org/10.1103/PhysRevE.95.053116}{doi:10.1103/PhysRevE.95.053116}).

\bibitem{ivers2017enumeration}
Ivers D. 2017  Enumeration, orthogonality and completeness of the
  incompressible {Coriolis} modes in a tri-axial ellipsoid. {\em Geophys.
  Astrophys. Fluid Dyn.} \textbf{111}, 333--354.
(\href{https://dx.doi.org/10.1080/03091929.2017.1330412}{doi:10.1080/03091929.2017.1330412}).

\bibitem{zhang2017theory}
Zhang K, Liao X. 2017 {\em Theory and modeling of rotating fluids: convection,
  inertial waves and precession}.
Cambridge, UK: Cambridge University Press.

\bibitem{papaloizou1978non}
Papaloizou J, Pringle JE. 1978  Non-radial oscillations of rotating stars and
  their relevance to the short-period oscillations of cataclysmic variables.
  {\em Mon. Not. R. Astron. Soc.} \textbf{182}, 423--442.
(\href{http://dx.doi.org/doi:10.1093/mnras/182.3.423}{doi:10.1093/mnras/182.3.423}).

\bibitem{lockitch1999r}
Lockitch KH, Friedman JL. 1999  Where are the r-modes of isentropic stars?.
  {\em Astrophys. J.} \textbf{521}, 764--788.
(\href{https://dx.doi.org/10.1086/307580}{doi:10.1086/307580}).

\bibitem{ivanov2010inertial}
Ivanov PB, Papaloizou JCB. 2010  Inertial waves in rotating bodies: a {WKBJ}
  formalism for inertial modes and a comparison with numerical results. {\em
  Mon. Not. R. Astron. Soc.} \textbf{407}, 1609--1630.
(\href{https://dx.doi.org/10.1111/j.1365-2966.2010.17009.x}{doi:10.1111/j.1365-2966.2010.17009.x}).

\bibitem{dintrans2001comparison}
Dintrans B, Rieutord M. 2001  A comparison of the anelastic and subseismic
  approximations for low-frequency gravity modes in stars. {\em Mon. Not. R.
  Astron. Soc.} \textbf{324}, 635--642.
(\href{https://dx.doi.org/10.1046/j.1365-8711.2001.04328.x}{doi:10.1046/j.1365-8711.2001.04328.x}).

\bibitem{wood2016oscillatory}
Wood TS, Bushby PJ. 2016  Oscillatory convection and limitations of the
  {Boussinesq} approximation. {\em J. Fluid Mech.} \textbf{803}, 502--515.
(\href{https://dx.doi.org/10.1017/jfm.2016.511}{doi:10.1017/jfm.2016.511}).

\bibitem{verhoeven2018validity}
Verhoeven J, Glatzmaier GA. 2018  Validity of sound-proof approaches in
  rapidly-rotating compressible convection: marginal stability versus
  turbulence. {\em Geophys. Astrophys. Fluid Dyn.} \textbf{112}, 36--61.
(\href{https://dx.doi.org/10.1080/03091929.2017.1380800}{doi:10.1080/03091929.2017.1380800}).

\bibitem{zhang2017shape}
Zhang K, Kong D, Schubert G. 2017  Shape, internal structure, zonal winds, and
  gravitational field of rapidly rotating {Jupiter}-like planets. {\em Annu.
  Rev. Earth Planet. Sci.} \textbf{45}, 419--446.
(\href{https://dx.doi.org/10.1146/annurev-earth-063016-020305}{doi:10.1146/annurev-earth-063016-020305}).

\bibitem{le2015flows}
Le~Bars M, C{\'e}bron D, Le~Gal P. 2015  Flows driven by libration, precession,
  and tides. {\em Annu. Rev. Fluid Mech.} \textbf{47}, 163--193.
(\href{https://dx.doi.org/10.1146/annurev-fluid-010814-014556}{doi:10.1146/annurev-fluid-010814-014556}).

\bibitem{davies2014strength}
Davies CJ, Stegman DR, Dumberry M. 2014  The strength of gravitational
  core-mantle coupling. {\em Geophys. Res. Lett.} \textbf{41}, 3786--3792.
(\href{https://dx.doi.org/10.1002/2014GL059836}{doi:10.1002/2014GL059836}).

\bibitem{cebron2014tidally}
C{\'e}bron D, Hollerbach R. 2014  Tidally driven dynamos in a rotating sphere.
  {\em Astrophys. J. Lett.} \textbf{789}, L25.
(\href{https://dx.doi.org/10.1088/2041-8205/789/1/L25}{doi:10.1088/2041-8205/789/1/L25}).

\bibitem{reddy2018turbulent}
Reddy KS, Favier B, Le~Bars M. 2018  Turbulent kinematic dynamos in ellipsoids
  driven by mechanical forcing. {\em Geophys. Res. Lett.} \textbf{45},
  1741--1750.
(\href{https://dx.doi.org/10.1002/2017GL076542}{doi:10.1002/2017GL076542}).

\bibitem{vidal2018magnetic}
Vidal J, C{\'e}bron D, Schaeffer N, Hollerbach R. 2018  Magnetic fields driven
  by tidal mixing in radiative stars. {\em Mon. Not. R. Astron. Soc.}
  \textbf{475}, 4579--4594.
(\href{https://dx.doi.org/10.1093/mnras/sty080}{doi:10.1093/mnras/sty080}).

\bibitem{herreman2011stokes}
Herreman W, Lesaffre P. 2011  Stokes drift dynamos. {\em J. Fluid Mech.}
  \textbf{679}, 32--57.
(\href{https://dx.doi.org/10.1017/jfm.2011.109}{doi:10.1017/jfm.2011.109}).

\bibitem{bryan1889waves}
Bryan GH. 1889  The waves on a rotating liquid spheroid of finite ellipticity.
  {\em Phil. Trans. R. Soc. A} \textbf{180}, 187--219.

\bibitem{rekier2018inertial}
Rekier J, Trinh A, Triana SA, Dehant V. 2018  Inertial modes in near-spherical
  geometries. {\em Geophys. J. Int.} \textbf{216}, 777--793.
(\href{https://dx.doi.org/10.1093/gji/ggy465}{doi:10.1093/gji/ggy465}).

\bibitem{reese2006acoustic}
Reese D, Ligni{\`e}res F, Rieutord M. 2006  Acoustic oscillations of rapidly
  rotating polytropic stars-{II}. {Effects} of the {Coriolis} and centrifugal
  accelerations. {\em Astron. Astrophys.} \textbf{455}, 621--637.
(\href{https://dx.doi.org/10.1051/0004-6361:20065269}{doi:10.1051/0004-6361:20065269}).

\bibitem{seyed2014dynamics}
Seyed-Mahmoud B, Moradi A. 2014  Dynamics of the {Earth}'s fluid core:
  {Implementation} of a {Clairaut} coordinate system. {\em Phys. Earth Planet.
  Int.} \textbf{227}, 61--67.
(\href{https://dx.doi.org/10.1016/j.pepi.2013.11.007}{doi:10.1016/j.pepi.2013.11.007}).

\bibitem{seyed2007inertial}
Seyed-Mahmoud B, Heikoop J, Seyed-Mahmoud R. 2007  Inertial modes of a
  compressible fluid core model. {\em Geophys. Astrophys. Fluid Dyn.}
  \textbf{101}, 489--505.
(\href{https://dx.doi.org/10.1080/03091920701523337}{doi:10.1080/03091920701523337}).

\bibitem{busse2005slow}
Busse F, Zhang K, Liao X. 2005  On slow inertial waves in the solar convection
  zone. {\em Astrophys. J.} \textbf{631}, L171.
(\href{https://dx.doi.org/10.1086/497300}{doi:10.1086/497300}).

\bibitem{wu2005origin}
Wu Y. 2005  Origin of tidal dissipation in {Jupiter}. {I}. {Properties} of
  inertial modes. {\em Astrophys. J.} \textbf{635}, 674.
(\href{https://dx.doi.org/10.1086/497354}{doi:10.1086/497354}).

\bibitem{clausen2014elliptical}
Clausen N, Tilgner A. 2014  Elliptical instability of compressible flow in
  ellipsoids. {\em Astron. Astrophys.} \textbf{562}, A25.
(\href{https://dx.doi.org/10.1051/0004-6361/201322817}{doi:10.1051/0004-6361/201322817}).

\bibitem{koulakis2018acoustic}
Koulakis JP, Pree S, Putterman S. 2018  Acoustic resonances in gas-filled
  spherical bulb with parabolic temperature profile. {\em J. Acoust. Soc. Am.}
  \textbf{144}, 2847--2851.
(\href{https://dx.doi.org/10.1121/1.5078599}{doi:10.1121/1.5078599}).

\bibitem{vantieghem2014inertial}
Vantieghem S. 2014  Inertial modes in a rotating triaxial ellipsoid. {\em Proc.
  R. Soc. A} \textbf{470}, 20140093.
(\href{https://dx.doi.org/10.1098/rspa.2014.0093}{doi:10.1098/rspa.2014.0093}).

\bibitem{vidal2020acoustics}
Vidal J, Su S, C{\'e}bron D. 2020  Compressible fluid modes in rigid
  ellipsoids: towards modal acoustic velocimetry. {\em J. Fluid Mech.}
  \textbf{885}, A39.
(\href{https://dx.doi.org/10.1017/jfm.2019.1004}{doi:10.1017/jfm.2019.1004}).

\bibitem{chandrasekhar1969ellipsoidal}
Chandrasekhar S. 1969 {\em Ellipsoidal figures of equilibrium}.
New York, USA: Dover Publications.

\bibitem{cowling1941non}
Cowling TG. 1941  The non-radial oscillations of polytropic stars. {\em Mon.
  Not. R. Astron. Soc.} \textbf{101}, 367.
(\href{https://dx.doi.org/10.1093/mnras/101.8.367}{doi:10.1093/mnras/101.8.367}).

\bibitem{cebron2012elliptical}
C{\'e}bron D, Le~Bars M, Moutou C, Le~Gal P. 2012  Elliptical instability in
  terrestrial planets and moons. {\em Astron. Astrophys.} \textbf{539}, A78.
(\href{https://dx.doi.org/10.1051/0004-6361/201117741}{doi:10.1051/0004-6361/201117741}).

\bibitem{abney1996ekman}
Abney M, Epstein RI. 1996  Ekman pumping in compact astrophysical bodies. {\em
  J. Fluid Mech.} \textbf{312}, 327--340.
(\href{https://dx.doi.org/10.1017/S0022112096002030}{doi:10.1017/S0022112096002030}).

\bibitem{glampedakis2006ekman}
Glampedakis K, Andersson N. 2006  Ekman layer damping of r modes revisited.
  {\em Mon. Not. R. Astron. Soc.} \textbf{371}, 1311--1321.
(\href{https://dx.doi.org/10.1111/j.1365-2966.2006.10749.x}{doi:10.1111/j.1365-2966.2006.10749.x}).

\bibitem{aldridge1969axisymmetric}
Aldridge KD, Toomre A. 1969  Axisymmetric inertial oscillations of a fluid in a
  rotating spherical container. {\em J. Fluid Mech.} \textbf{37}, 307--323.
(\href{https://dx.doi.org/10.1017/S0022112069000565}{doi:10.1017/S0022112069000565}).

\bibitem{roberts1967introduction}
Roberts PH. 1967 {\em An {Introduction} to {Magnetohydrodynamics}}.
New York, USA: Longmans.

\bibitem{braviner2014tidal}
Braviner HJ, Ogilvie GI. 2014  Tidal interactions of a {Maclaurin}
  spheroid--{I}. {Properties} of free oscillation modes. {\em Mon. Not. R.
  Astron. Soc.} \textbf{441}, 2321--2345.
(\href{https://dx.doi.org/10.1093/mnras/stu704}{doi:10.1093/mnras/stu704}).

\bibitem{goodman2009dynamical}
Goodman J, Lackner C. 2009  Dynamical tides in rotating planets and stars. {\em
  Astrophys. J.} \textbf{696}, 2054.
(\href{https://dx.doi.org/10.1088/0004-637X/696/2/2054}{doi:10.1088/0004-637X/696/2/2054}).

\bibitem{kapyla2017convection}
K{\"a}pyl{\"a} PJ, K{\"a}pyl{\"a} MJ, Olspert N, Warnecke J, Brandenburg A.
  2017  Convection-driven spherical shell dynamos at varying {Prandtl} numbers.
  {\em Astron. Astrophys.} \textbf{599}, A4.
(\href{https://dx.doi.org/10.1051/0004-6361/201628973}{doi:10.1051/0004-6361/201628973}).

\bibitem{liu2019onset}
Liu S, Wan ZH, Yan R, Sun C, Sun DJ. 2019  Onset of fully compressible
  convection in a rapidly rotating spherical shell. {\em J. Fluid Mech.}
  \textbf{873}, 1090--1115.
(\href{https://dx.doi.org/10.1017/jfm.2019.436}{doi:10.1017/jfm.2019.436}).

\bibitem{labrosse2015thermal}
Labrosse S. 2015  Thermal evolution of the core with a high thermal
  conductivity. {\em Phys. Earth Planet. Inter.} \textbf{247}, 36--55.
(\href{https://dx.doi.org/10.1016/j.pepi.2015.02.002}{doi:10.1016/j.pepi.2015.02.002}).

\bibitem{evonuk2012simulating}
Evonuk M, Samuel H. 2012  Simulating rotating fluid bodies: {When} is vorticity
  generation via density-stratification important?. {\em Earth Planet. Sc.
  Lett.} \textbf{317}, 1--7.
(\href{https://dx.doi.org/10.1016/j.epsl.2011.11.036}{doi:10.1016/j.epsl.2011.11.036}).

\bibitem{gerick2020mhd}
Gerick F, Jault D, Noir J, Vidal J. 2020  Pressure torque of torsional
  {Alfv\'en} modes acting on an ellipsoidal mantle. {\em Geophys. J. Int.}
  \textbf{222}, 338--351.
(\href{https://dx.doi.org/10.1093/gji/ggaa166}{doi:10.1093/gji/ggaa166}).

\bibitem{chandrasekhar1958introduction}
Chandrasekhar S. 1958 {\em An introduction to the study of stellar structure}.
New York, USA: Dover Publications.

\bibitem{lai1993ellipsoidal}
Lai D, Rasio FA, Shapiro SL. 1993  Ellipsoidal figures of equilibrium:
  {Compressible} models. {\em Astrophys. J. Suppl. S.} \textbf{88}, 205--252.
(\href{https://dx.doi.org/10.1086/191822}{doi:10.1086/191822}).

\bibitem{lynden1967stability}
Lynden-Bell D, Ostriker JP. 1967  On the stability of differentially rotating
  bodies. {\em Mon. Not. R. Astron. Soc.} \textbf{136}, 293--310.
(\href{https://dx.doi.org/10.1093/mnras/136.3.293}{doi:10.1093/mnras/136.3.293}).

\bibitem{valette1989spectre}
Valette B. 1989  Spectre des vibrations propres d'un corps {\'e}lastique,
  auto-gravitant, en rotation uniforme et contenant une partie fluide. {\em C.
  R. Acad. Sci. Paris} \textbf{309}, 419--422.

\bibitem{barston1967eigenvalue}
Barston EM. 1967  Eigenvalue problem for {Lagrangian} systems. {\em J. Math.
  Phys.} \textbf{8}, 523--532.
(\href{https://dx.doi.org/10.1063/1.1705227}{doi:10.1063/1.1705227}).

\bibitem{lebovitz1989stability}
Lebovitz NR. 1989  The stability equations for rotating, inviscid fluids:
  {Galerkin} methods and orthogonal bases. {\em Geophys. Astrophys. Fluid Dyn.}
  \textbf{46}, 221--243.
(\href{https://dx.doi.org/10.1080/03091928908208913}{doi:10.1080/03091928908208913}).

\bibitem{sobouti1981potentials}
Sobouti Y. 1981  The potentials for the g-,p- and the toroidal modes of
  self-gravitating fluids. {\em Astron. Astrophys.} \textbf{100}, 319--322.

\bibitem{tisseur2001quadratic}
Tisseur F, Meerbergen K. 2001  The quadratic eigenvalue problem. {\em SIAM
  Rev.} \textbf{43}, 235--286.
(\href{https://dx.doi.org/10.1137/S0036144500381988}{doi:10.1137/S0036144500381988}).

\bibitem{chaljub2004spectral}
Chaljub E, Valette B. 2004  Spectral element modelling of three-dimensional
  wave propagation in a self-gravitating {Earth} with an arbitrarily stratified
  outer core. {\em Geophys. J. Int.} \textbf{158}, 131--141.
(\href{https://dx.doi.org/10.1111/j.1365-246X.2004.02267.x}{doi:10.1111/j.1365-246X.2004.02267.x}).

\bibitem{dyson1979perturbations}
Dyson J, Schutz BF. 1979  Perturbations and stability of rotating stars. {I}.
  {Completeness} of normal modes. {\em Proc. R. Soc. Lond. A} \textbf{368},
  389--410.
(\href{https://dx.doi.org/10.1098/rspa.1979.0137}{doi:10.1098/rspa.1979.0137}).

\bibitem{su2020zoro}
Su S, C{\'e}bron D, Nataf HC, Cardin P, Vidal J, Solazzo M, Do Y. 2020
  Acoustic spectra of a gas-filled rotating spheroid. {\em Eur. J. Mech.
  B-Fluid.} \textbf{84}, 302--310.
(\href{https://doi.org/10.1016/j.euromechflu.2020.03.003}{doi:10.1016/j.euromechflu.2020.03.003}).

\bibitem{backus1961rotational}
Backus G, Gilbert F. 1961  The rotational splitting of the free oscillations of
  the {Earth}. {\em Proc. Natl Acad. Sci. USA} \textbf{47}, 362--371.
(\href{https://dx.doi.org/10.1073/pnas.47.3.362}{doi:10.1073/pnas.47.3.362}).

\bibitem{maffei2017characterization}
Maffei S, Jackson A, Livermore PW. 2017  Characterization of columnar inertial
  modes in rapidly rotating spheres and spheroids. {\em Proc. R. Soc. A}
  \textbf{473}, 20170181.
(\href{https://dx.doi.org/10.1098/rspa.2017.0181}{doi:10.1098/rspa.2017.0181}).

\bibitem{valette1989etude}
Valette B. 1989  {\'E}tude d'une classe de probl{\`e}mes spectraux. {\em C. R.
  Acad. Sci. Paris} \textbf{309}, 785--788.

\bibitem{bardsley2018could}
Bardsley OP. 2018  Could hydrodynamic {Rossby} waves explain the westward
  drift?. {\em Proc. R. Soc. A} \textbf{474}, 20180119.
(\href{https://dx.doi.org/10.1098/rspa.2018.0119}{doi:10.1098/rspa.2018.0119}).

\bibitem{ivanov2013unified}
Ivanov PB, Papaloizou JCB, Chernov SV. 2013  A unified normal mode approach to
  dynamic tides and its application to rotating {Sun}-like stars. {\em Mon.
  Not. R. Astron. Soc.} \textbf{432}, 2339--2365.
(\href{https://dx.doi.org/10.1093/mnras/stt595}{doi:10.1093/mnras/stt595}).

\bibitem{schutz1978langrangian}
Schutz BF, Friedman JL. 1978  Langrangian perturbation theory of
  nonrelativistic fluids. {\em Astrophys. J.} \textbf{221}, 937--957.
(\href{https://dx.doi.org/10.1086/156098}{doi:10.1086/156098}).

\bibitem{vantieghem2015latitudinal}
Vantieghem S, C{\'e}bron D, Noir J. 2015  Latitudinal libration driven flows in
  triaxial ellipsoids. {\em J. Fluid Mech.} \textbf{771}, 193--228.
(\href{https://dx.doi.org/10.1017/jfm.2015.130}{doi:10.1017/jfm.2015.130}).

\bibitem{wahr1981normal}
Wahr JM. 1981  A normal mode expansion for the forced response of a rotating
  {Earth}. {\em Geophys. J. Int.} \textbf{64}, 651--675.
(\href{https://dx.doi.org/10.1111/j.1365-246X.1981.tb02689.x}{doi:10.1111/j.1365-246X.1981.tb02689.x}).

\bibitem{schutz1980perturbations}
Schutz BF. 1980  Perturbations and stability of rotating stars--{II}.
  {Properties} of the eigenvectors and a variational principle. {\em Mon. Not.
  R. Astron. Soc.} \textbf{190}, 7--20.
(\href{https://dx.doi.org/10.1093/mnras/190.1.7}{doi:10.1093/mnras/190.1.7}).

\bibitem{greenspan1969non}
Greenspan HP. 1969  On the non-linear interaction of inertial modes. {\em J.
  Fluid Mech.} \textbf{36}, 257--264.
(\href{https://dx.doi.org/10.1017/S0022112069001649}{doi:10.1017/S0022112069001649}).

\bibitem{busse1976simple}
Busse FH. 1976  A simple model of convection in the {Jovian} atmosphere. {\em
  Icarus} \textbf{29}, 255--260.
(\href{https://doi.org/10.1016/0019-1035(76)90053-1}{doi:10.1016/0019-1035(76)90053-1}).

\bibitem{glatzmaier2018computer}
Glatzmaier GA. 2018  Computer simulations of {Jupiter's} deep internal dynamics
  help interpret what {Juno} sees. {\em Proc. Natl Acad. Sci. USA}
  \textbf{115}, 6896--6904.
(\href{https://dx.doi.org/10.1073/pnas.1709125115}{doi:10.1073/pnas.1709125115}).

\bibitem{guillot2018suppression}
Guillot T et~al.. 2018  A suppression of differential rotation in {Jupiter}'s
  deep interior. {\em Nature} \textbf{555}, 227--230.
(\href{https://dx.doi.org/10.1038/nature25775}{doi:10.1038/nature25775}).

\bibitem{moore2019time}
Moore KM, Cao H, Bloxham J, Stevenson DJ, Connerney JEP, Bolton SJ. 2019  Time
  variation of {Jupiter's} internal magnetic field consistent with zonal wind
  advection. {\em Nat. Astron.} \textbf{3}, 730--735.
(\href{https://dx.doi.org/10.1038/s41550-019-0772-5}{doi:10.1038/s41550-019-0772-5}).

\bibitem{christensen2020mechanisms}
Christensen UR, Wicht J, Dietrich W. 2020  Mechanisms for limiting the depth of
  zonal winds in the gas giant planets. {\em Astrophys. J.} \textbf{890}, 61.
(\href{https://dx.doi.org/10.3847/1538-4357/ab698c}{doi:10.3847/1538-4357/ab698c}).

\bibitem{lebovitz1996new}
Lebovitz NR, Lifschitz A. 1996  New global instabilities of the {Riemann}
  ellipsoids. {\em Astrophys. J.} \textbf{458}, 699.
(\href{https://dx.doi.org/10.1086/176851}{doi:10.1086/176851}).

\bibitem{vidal2017inviscid}
Vidal J, C{\'e}bron D. 2017  Inviscid instabilities in rotating ellipsoids on
  eccentric {Kepler} orbits. {\em J. Fluid Mech.} \textbf{833}, 469--511.
(\href{https://dx.doi.org/10.1017/jfm.2017.689}{doi:10.1017/jfm.2017.689}).

\bibitem{stewartson1963motion}
Stewartson K, Roberts PH. 1963  On the motion of liquid in a spheroidal cavity
  of a precessing rigid body. {\em J.Fluid Mech.} \textbf{17}, 1--20.
(\href{https://dx.doi.org/10.1017/S0022112063001063}{doi:10.1017/S0022112063001063}).

\bibitem{lin2015shear}
Lin Y, Marti P, Noir J. 2015  Shear-driven parametric instability in a
  precessing sphere. {\em Phys. Fluids} \textbf{27}, 046601.

\bibitem{hollerbach1995oscillatory}
Hollerbach R, Kerswell RR. 1995  Oscillatory internal shear layers in rotating
  and precessing flows. {\em J. Fluid Mech.} \textbf{298}, 327--339.
(\href{https://dx.doi.org/10.1017/S0022112095003338}{doi:10.1017/S0022112095003338}).

\bibitem{liao2001viscous}
Liao X, Zhang K, Earnshaw P. 2001  On the viscous damping of inertial
  oscillation in planetary fluid interiors. {\em Phys. Earth Planet. Inter.}
  \textbf{128}, 125--136.
(\href{https://dx.doi.org/10.1016/S0031-9201(01)00281-3}{doi:10.1016/S0031-9201(01)00281-3}).

\bibitem{liao2010asymptotic}
Liao X, Zhang K. 2010  Asymptotic and numerical solutions of the initial value
  problem in rotating planetary fluid cores. {\em Geophys. J. Int.}
  \textbf{180}, 181--192.
(\href{https://dx.doi.org/10.1111/j.1365-246X.2009.04421.x}{doi:10.1111/j.1365-246X.2009.04421.x}).

\bibitem{lemasquerier2017libration}
Lemasquerier D, Grannan AM, Vidal J, C{\'e}bron D, Favier B, Le~Bars M, Aurnou
  JM. 2017  Libration-driven flows in ellipsoidal shells. {\em J. Geophys. Res.
  Planets} \textbf{122}, 1926--1950.
(\href{https://dx.doi.org/10.1002/2017JE005340}{doi:10.1002/2017JE005340}).

\bibitem{lebovitz1979ellipsoidal}
Lebovitz NR. 1979  Ellipsoidal potentials of polynomial distributions of
  matter. {\em Astrophys. J.} \textbf{234}, 619--627.
(\href{https://dx.doi.org/10.1086/157538}{doi:10.1086/157538}).

\bibitem{vidal2019acoustics}
Vidal J, Su S, C\'ebron D. 2019  Polynomial description of acoustic modes in
  fluid ellipsoids. In {\em Comptes-Rendus de la 22e Rencontre du
  Non-Lin{\'e}aire}.
(\href{https://hal.archives-ouvertes.fr/hal-02200485}{hal-02200485}).

\bibitem{kerswell2002elliptical}
Kerswell RR. 2002  Elliptical instability. {\em Annu. Rev. Fluid Mech.}
  \textbf{34}, 83--113.
(\href{https://dx.doi.org/10.1146/annurev.fluid.34.081701.171829}{doi:10.1146/annurev.fluid.34.081701.171829}).

\bibitem{cebron2013elliptical}
C{\'e}bron D, Le~Bars M, Le~Gal P, Moutou C, Leconte J, Sauret A. 2013
  Elliptical instability in hot {Jupiter} systems. {\em Icarus} \textbf{226},
  1642--1653.
(\href{https://dx.doi.org/10.1016/j.icarus.2012.12.017}{doi:10.1016/j.icarus.2012.12.017}).

\bibitem{lignieres2009asymptotic}
Ligni{\`e}res F, Georgeot B. 2009  Asymptotic analysis of high-frequency
  acoustic modes in rapidly rotating stars. {\em Astron. Astrophys.}
  \textbf{500}, 1173--1192.
(\href{https://dx.doi.org/10.1051/0004-6361/200811165}{doi:10.1051/0004-6361/200811165}).

\bibitem{prat2016asymptotic}
Prat V, Ligni{\`e}res F, Ballot J. 2016  Asymptotic theory of gravity modes in
  rotating stars-{I}. {Ray} dynamics. {\em Astron. Astrophys.} \textbf{587},
  A110.
(\href{https://dx.doi.org/10.1051/0004-6361/201527737}{doi:10.1051/0004-6361/201527737}).

\bibitem{friedlander1982internal}
Friedlander S, Siegmann WL. 1982  Internal waves in a rotating stratified fluid
  in an arbitrary gravitational field. {\em Geophys. Astrophys. Fluid Dyn.}
  \textbf{19}, 267--291.
(\href{https://dx.doi.org/10.1080/03091928208208959}{doi:10.1080/03091928208208959}).

\bibitem{seyed2015effects}
Seyed-Mahmoud B, Moradi A, Kamruzzaman M, Naseri H. 2015  Effects of density
  stratification on the frequencies of the inertial-gravity modes of the
  {Earth}'s fluid core. {\em Geophys. J. Int.} \textbf{202}, 1146--1157.
(\href{https://dx.doi.org/10.1093/gji/ggv215}{doi:10.1093/gji/ggv215}).

\bibitem{kerswell1993elliptical}
Kerswell RR. 1993  Elliptical instabilities of stratified, hydromagnetic waves.
  {\em Geophys. Astrophys. Fluid Dyn.} \textbf{71}, 105--143.
(\href{https://dx.doi.org/10.1080/03091929308203599}{doi:10.1080/03091929308203599}).

\bibitem{vidal2019binaries}
Vidal J, C{\'e}bron D, ud~Doula A, Alecian E. 2019  Fossil field decay due to
  nonlinear tides in massive binaries. {\em Astron. Astrophys.} \textbf{629},
  A142.
(\href{https://dx.doi.org/10.1051/0004-6361/201935658}{doi:10.1051/0004-6361/201935658}).

\bibitem{schaeffer2013efficient}
Schaeffer N. 2013  Efficient spherical harmonic transforms aimed at
  pseudospectral numerical simulations. {\em Geochem. Geophy. Geos.}
  \textbf{14}, 751--758.
(\href{https://dx.doi.org/10.1002/ggge.20071}{doi:10.1002/ggge.20071}).

\bibitem{cartan1922petites}
Cartan ME. 1922  Sur les petites oscillations d'une masse de fluide. {\em Bull.
  Sci. Math.} \textbf{46}, 317--369.

\bibitem{guermond2013remarks}
Guermond JL, L{\'e}orat J, Luddens F, Nore C. 2013  Remarks on the stability of
  the {Navier}--{Stokes} equations supplemented with stress boundary
  conditions. {\em Eur. J. Mech. B-Fluid} \textbf{39}, 1--10.
(\href{https://dx.doi.org/10.1016/j.euromechflu.2012.11.003}{doi:10.1016/j.euromechflu.2012.11.003}).

\end{thebibliography}
\bibliographystyle{RS.bst}
}

\end{document}